# Who is Responsible When AI Fails? Mapping Causes, Entities, and Consequences of AI Privacy and Ethical Incidents

**Hilda Hadan, Reza Hadi Mogavi, Leah Zhang-Kennedy & Lennart E. Nacke**

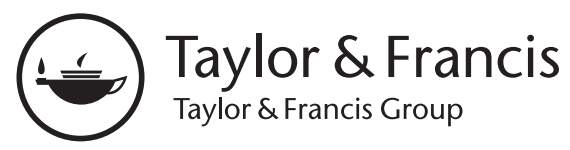

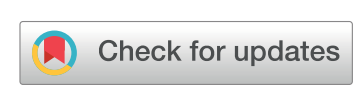

# Who is Responsible When AI Fails? Mapping Causes, Entities, and Consequences of AI Privacy and Ethical Incidents

Hilda Hadan, Reza Hadi Mogavi, Leah Zhang-Kennedy and Lennart E. Nacke

Stratford School of Interaction Design and Business, University of Waterloo, Stratford, ON, Canada

**ABSTRACT**
The rapid growth of artificial intelligence (AI) technologies has raised major privacy and ethical concerns. However, existing AI incident taxonomies and guidelines lack grounding in real-world cases, limiting their effectiveness for prevention and mitigation. We analyzed 202 real-world AI privacy and ethical incidents to develop a taxonomy that classifies them across AI lifecycle stages and captures contributing factors, including causes, responsible entities, sources of disclosure, and impacts. Our findings reveal widespread harms from poor organizational decisions and legal non-compliance, limited corrective interventions, and rare reporting from AI developers and adopting entities. Our taxonomy offers a structured approach for systematic incident reporting and emphasizes the weaknesses of current AI governance frameworks. Our findings provide actionable guidance for policymakers and practitioners to strengthen user protections, develop targeted AI policies, enhance reporting practices, and foster responsible AI governance and innovation, especially in contexts such as social media and child protection.

## 1. Introduction

In recent years, a leading Artificial Intelligence (AI) chatbot exposed users' private chat histories and sensitive personal data during its conversations with others (AIAAIC Repository, 2023a). A few months later, an AI-powered companion posed as a 25-year-old man and lured a 13-year-old girl to a nearby park (AIAAIC Repository, 2025a). While AI is transforming human decision-making and driving innovation in fields such as art, education, and games (Capel & Brereton, 2023; Hadan et al., 2024; Mogavi et al., 2024), these incidents demonstrate how AI can also pose serious threats to human privacy and vulnerable populations. Public opinion on AI is controversial, with a global movement of scientists and industry leaders calling for a "*pause*" in AI deployment (Stichting PauseAI, 2024). A series of AI Summits with government representatives was also held to discuss the global response to AI risks (AI Action Summit, 2025). The core concerns include AI privacy violations (Lee et al., 2024; Shelby et al., 2023; Slattery et al., 2024), the reinforcement of bias and discrimination (Mogavi et al., 2024; Weidinger et al., 2022), and misuse for deepfakes, harassment, and scams (Lee et al., 2024; Stichting PauseAI, 2024). The rapid pace of AI innovation and the growing number of related incidents emphasize the need for effective governance and reliable risk prevention and mitigation strategies. However, the complexity of AI behaviors often hampers practitioners from implementing these measures prior to deployment (Holstein et al., 2019; Turri & Dzombak, 2023).

To navigate the complexity of AI incidents and foster effective AI governance and incident prevention, our research develops a taxonomy based on a thematic analysis of $N = 202$ real-world AI privacy and ethical incidents from the AIAAIC repository. The research question (RQ) guiding our study is: **What are the common types of AI incidents and their contributing contextual factors across the stages of AI system lifecycle?** While previous efforts from academia, industry, and government have proposed taxonomies and guidelines for AI incidents and risks, they often lack grounding in real-world incidents (e.g., Dafoe, 2024; Gov.UK, 2024; Zeng et al., 2024), focus exclusively on incident types

**CONTACT** Hilda Hadan hhadan@uwaterloo.ca Stratford School of Interaction Design and Business, University of Waterloo, 125 St Patrick St, Stratford, ON N5A 0C1, Canada without considering contextual factors (e.g., Lee et al., 2024; Shahriar et al., 2023; Weidinger et al., 2022), or are limited to AI's application in specific domains (e.g., Golpayegani et al., 2022; Hutiri et al., 2024; Shahriar et al., 2023). As a result, these taxonomies cannot inform effective AI incident governance and prevention because they lack actionable insights into the factors that contribute to AI incidents.

Our research provides a detailed taxonomy that classifies AI incident types across various stages of the AI lifecycle, and contributing factors such as causes, responsible entities, sources of disclosure, and consequential impacts. We developed this taxonomy through a thematic analysis of 202 real-world AI privacy and ethical incidents from the AIAAIC repository, the largest and most up-to-date crowdsourced collection of AI incidents (Lee et al., 2024; Raji et al., 2022; Turri & Dzombak, 2023). Our methodology follows recommendations from prior AI literature (Turri & Dzombak, 2023; Velázquez et al., 2024) and builds on structured approaches from cybersecurity incidents and domain-specific AI incidents analyses (e.g., Golpayegani et al., 2022; Hadan et al., 2021; Hutiri et al., 2024; Serrano et al., 2019). To ensure that our taxonomy is both comprehensive and proactive in preventing current and future incidents (Turri & Dzombak, 2023), we analyzed cases involving confirmed harms as well as those posing risks under government investigation or subject to public criticism. Our study focuses on AI privacy and ethical incidents, which are frequently reported and highly impactful (Golpayegani et al., 2022; Slattery et al., 2024). We analyze these incidents in a same taxonomy, as privacy and ethical violations are closely intertwined, with ethical violations often leading to privacy harms (Weidinger et al., 2022) and privacy is an essential principle of ethical AI development (Jobin et al., 2019).

Our research advances the understanding of AI incidents and offers actionable insights for policymakers and practitioners in AI, privacy, and ethics to improve AI governance, incident detection, and risk mitigation. Our findings revealed the lack of incident reporting by AI developers and adopting organizations and government entities, the prevalence of incidents caused by organizational decisions and legal non-compliance, and the limited number of incidents resulting in legal actions, corrective measures, or risk-mitigation interventions. Specifically, we make four main contributions:

1. We present an empirically-derived taxonomy of AI privacy and ethical incidents and their contributing factors based on analysis of 202 real-world cases from 2023–2024.
2. Our findings identify gaps in current AI governance frameworks through empirical evidence, which show critical areas of organizational non-compliance and insufficient oversight.
3. Our research offers evidence-based recommendations for improving AI incident reporting and prevention, including specific mechanisms for mandatory disclosure, enhanced monitoring systems, and platform-specific content moderation policies.
4. Through a comparison, we demonstrate that our taxonomy captures both previously hypothesized AI incidents identified in the literature (Lee et al., 2024; Shelby et al., 2023; Slattery et al., 2024) and novel incidents and contextual factors outside their scope.

Overall, our findings suggest that current AI governance frameworks are inadequate, and we urgently need stronger user protections and targeted AI policies to address the widespread and largely unregulated harms of AI, especially on social media and among children. As industrial and governmental entities begin proposing risk management frameworks to ensure ethical AI development and deployment (e.g., OECD, n.d; European Commission, 2019; The White House, 2020; NIST, 2023), our taxonomy and findings can inform the ethical and safe design and implementation of AI systems, and enhance AI governance, incident detection, and risk mitigation.

## 2. Related work

In this section, we summarize the emerging risks, concerns, and threats associated with the evolution of AI, as well as the current AI incident taxonomies proposed by academia, private organizations, and government authorities. Finally, we demonstrate how our research addresses gaps in these existing taxonomies and suggest enhancements for systematic AI incident reporting, and propose recommendations for regulating and preventing AI incidents in the future.

### 2.1. AI and its threats, risks, and concerns

As AI technologies advance, related risks and concerns have also emerged. AI brought benefits in various areas (Capel & Brereton, 2023; Hadan et al., 2024; Mogavi et al., 2024) but introduced harms through its limitations in algorithmic accuracy, high resource demands, and biases inherent in training data or arising from human misuse (Nah et al., 2023; Slattery et al., 2024; Weidinger et al., 2022).

#### 2.1.1. Reinforcement of discrimination, hate speech, and exclusion bias

As AI models are trained on online sources, they tend to replicate social norms embedded in their training data (Mogavi et al., 2024; Weidinger et al., 2022). Consequently, these models may reinforce stereotypes and inequalities (Caliskan et al., 2017; Nadeem et al., 2021; Rae et al., 2021), and exclude historically marginalized social groups (Bender et al., 2021; Keyes et al., 2021; Rincón et al., 2021). For example, an AI might define "family" narrowly as a married heterosexual couple with biological children, excluding diverse family structures (Weidinger et al., 2022). AI may wrongly reject loan or mortgage applications (Shelby et al., 2023; Wirtz et al., 2022), discriminate against qualified candidates (Kamila & Jasrotia, 2025; Shelby et al., 2023; Wirtz et al., 2022), or misidentify and wrongfully arrest innocent people (Kamila & Jasrotia, 2025; Paes et al., 2023). Even worse, AI can internalize and reproduce hate speech, offensive language, or violent incitement found online (Fortuna & Nunes, 2019; Gorwa et al., 2020; Shelby et al., 2023). The lack of transparency in such systems further exacerbates the issue, making it challenging for affected people to seek justice (Vredenburgh, 2022). These biases persist across cultures (Mogavi et al., 2024) and are often exacerbated in underrepresented languages, dialects, or sociolects (Joshi et al., 2020; Nozza et al., 2021), causing AI models to perform inconsistently and amplifying existing inequalities (Weidinger et al., 2022; Welbl et al., 2021). Furthermore, AI models trained on data from specific time periods may reinforce outdated norms and cannot reflect evolving social contexts (Bender et al., 2021; Joshi et al., 2020). Furthermore, if AI models are trained only on data from a specific time and social context, they risk "locking in" outdated norms and might not reflect social changes (Haraway, 2004; Weidinger et al., 2022).

#### 2.1.2. Privacy risks and information hazards from AI inferences

The vast amounts of data needed to support and improve AI model performance raise privacy concerns, as much of this data originates from people (Lee et al., 2024; Radanliev, 2025; Wach et al., 2023). Accidental leaks can occur when AI models inadvertently "remember" and reproduce personal information from their training datasets, such as names, addresses, or medical records (Shelby et al., 2023; Slattery et al., 2024). Advanced AI models also have the capability to infer and reveal sensitive information not directly included in their training data or user inputs (Cunha & Estima, 2023; Weidinger et al., 2022). Literature believed that the reveal of sensitive information by AI can be similar to "doxing" and potentially cause psychological and material harm (Douglas, 2016; Weidinger et al., 2022), with studies calling for protecting user data through AI algorithm design (Radanliev, 2025). In fact, public datasets and language analysis are increasingly used to infer personal traits (Kosinski et al., 2013; Park et al., 2015; Quercia et al., 2011), despite ethical concerns (Vicent, 2017). For example, data from social media platforms has been used to predict political orientation (Makazhanov & Rafiei, 2013; Mao et al., 2011; Preoţiuc-Pietro et al., 2017), age (Morgan-Lopez et al., 2017; Nguyen et al., 2013), and health conditions (Golbeck, 2018). Even when such AI-generated inferences are inaccurate, they can still result in discrimination and harm if treated as correct (Weidinger et al., 2022). Furthermore, AI-powered mass surveillance can become more cost-effective and widespread, potentially enabling illegitimate censorship and abuses of privacy and democratic rights (Lee et al., 2024).

#### 2.1.3. Misinformation from AI limitations

AI models, which are designed to predict likely sequences of words, may unintentionally generate and spread factually inaccurate information when trained on false or outdated data (Weidinger et al., 2022). The accuracy of a statement often depends on contextual factors such as location, timing, or the speaker's mental and social circumstances, which are typically missing from training datasets and cannot be interpreted by AI models (Bender & Koller, 2020). This limitation, known as the "symbol grounding problem," refers to AI's limitation in connecting abstract symbols (like words) with their real-world meanings and contexts (Harnad, 1990; Weidinger et al., 2022). Consequently, AI-generated misinformation can reinforce false beliefs, threaten user autonomy by promoting misconceptions (Lin et al., 2022), marginalize minority views as incorrect (Weidinger et al., 2022), and erode trust in reliable information sources (Ognyanova et al., 2020). In critical fields like medicine and law, such inaccuracies are particularly dangerous, as incorrect medical advice can harm patients (Miner et al., 2016), and false legal advice may lead to poor decisions (Bikeev et al., 2019). Even in less sensitive areas, misinformation can cause actions users might otherwise avoid (Weidinger et al., 2022).

#### 2.1.4. Disinformation and cyberattacks augmented by AI

AI models can be intentionally deployed to generate large volumes of targeted disinformation, misleading the public, influencing opinions, manipulating stock market prices, or creating a false "majority opinion" by flooding online platforms with AI-generated text (Weidinger et al., 2022). Research indicates that people often struggle to distinguish between human-written and AI-generated content (Hadan et al., 2024; Longoni et al., 2022). As a result, AI can be used to enhance scams, impersonate communication styles for identity theft and academic cheating, and personalize scam emails to improve their effectiveness (Weidinger et al., 2022). Malicious actors can craft inputs that exploit a model's knowledge of sensitive information (Cui et al., 2024). Additionally, AI-powered code generation tools can also create advanced malware that avoids detection (Zeng et al., 2024). AI-generated disinformation in defense systems may also distract security experts from real threats (Ranade et al., 2021). These risks will increase as AI systems become more accessible (Weidinger et al., 2022).

#### 2.1.5. Environmental costs and socioeconomic disruptions due to AI automation

Like many advanced technologies, AI also has significant environmental and socioeconomic impacts. Training and operating AI models require substantial energy, leading to high carbon emissions and consumption of fresh water for cooling data centers, which can affect ecosystems (Bender et al., 2021; Rae et al., 2021; Strubell et al., 2020; Weidinger et al., 2022). AI's capability to automate tasks may result in job displacement for low-skilled workers and shift job types to low-wage roles (Farahani & Ghasemi, 2024; Georgieff & Milanez, 2021; Moradi & Levy, 2020; Webb, 2019). AI-powered applications can also reduce job quality by speeding up task completion, increasing work pace, and diminishing worker autonomy and satisfaction (Green et al., 2023; Moradi & Levy, 2020; Spencer, 2025). Creative industries also face disruption as AI models emulate artistic styles without strictly infringing copyright, potentially reducing demand for original creative work and impacting artists' livelihoods (Hadan et al., 2024; Köbis & Mossink, 2021; Mogavi et al., 2024). Moreover, access to AI and its benefits may be unevenly distributed due to differences in hardware availability, skill levels, or internet access, potentially widening economic inequalities and primarily benefiting wealthier and technologically advanced groups (Bender et al., 2021; Capraro et al., 2024).

### 2.2. Existing AI incident and risk taxonomies

Many taxonomies have been developed for classifying incidents and risks associated with AI systems. Early work by Yampolskiy (2016) proposed a taxonomy grouping AI risks based on when they happen (such as before or after deployment), their causes (such as human actions, poor design, or environmental factors), and unexpected risks from the AI system's self-improvement Weidinger et al. (2022) introduced a taxonomy focused on risks from language models. This taxonomy included issues like discrimination, hate speech, exclusion, information hazards, misinformation, harmful uses, problems in human–computer interaction, and environmental or social impacts (Weidinger et al., 2022, 2023). This taxonomy is often used in AI research because it covers a wide range of relevant concerns (e.g., Achiam et al., 2023; Shelby et al., 2023; Slattery et al., 2024). Building on these earlier works, Slattery et al. (2024) combined different taxonomies into two unified frameworks. These frameworks describe AI risks by looking at who is responsible (e.g., a human, an AI system, or another party), when the risk occurs (before or after deployment), whether the harm is intentional or accidental, and the type of harm involved (Slattery et al., 2024). Similarly, Velázquez et al. (2024) developed a taxonomy based on real AI incidents. Their framework classifies incidents by type of harm, the source of the problem, the people or groups affected, and moral judgments. In the regulatory domain, Golpayegani et al. (2023) created a taxonomy based on the European Union's AI Act. Zeng et al. (2024) expanded on this by studying policies from leading corporate policies and AI-related government regulations, highlighting AI systems' operational risks, social and legal issues, and human rights concerns.

Other taxonomies focus on specific AI systems or areas. In the privacy domain, Shahriar et al. (2023) identified four main types of AI privacy risks: identification, inaccurate decisions, lack of transparency, and non-compliance with laws. Their taxonomy also suggests ways to reduce these risks across different stages of the AI lifecycle, offering useful guidance for addressing privacy issues (Shahriar et al., 2023). Building on Solove et al.'s (2006) well-known privacy risk taxonomy for traditional technologies, Lee et al. (2024) analyzed the real-world AI privacy incidents and created a taxonomy that shows how AI has made existing privacy problems worse and has also introduced new issues. In the financial sector, Giudici and Raffinetti (2023) developed a taxonomy for AI-related risks in regulation and also proposed a model to measure these risks in financial systems. In public health, Golpayegani et al. (2022) created a taxonomy that includes not just types of risks but also important context, including the purpose of the AI, the people involved, and the negative effects of incidents. Frameworks have also been introduced to classify AI-resulted harms and ethical issues. For example, scholars have created taxonomies of socio-technical harms (Shelby et al., 2023), technical issues that lead to AI risks (Cummings, 2024), and general harms like threats to personal freedom, safety, and reputation (Abercrombie et al., 2024). Ethical and safety issues related to AI speech systems have also been explored (Hutiri et al., 2024). Additionally, Burema et al. (2023) studied AI incidents across sectors (e.g., policing, education) and found that many ethical concerns are shaped by the unique structures and practices of each sector.

Governments and organizations have also created taxonomies and tools to help report AI incidents, manage risks, and design ethical AI systems. For example, the Organisation for Economic Co-operation and Development (OECD) created a framework to evaluate AI systems using five areas: people and the environment, economic context, data inputs, AI model, and outputs. This framework helps assess risks and check if AI systems follow OECD's values and principles (OECD, 2023). The Center for Security and Emerging Technology (CSET) introduced the AI Harm Framework, which separates harms into two types (i.e., tangible and intangible) and two stages (i.e., potential and actual) and offers flexible tools for different situations (Hoffmann & Frase, 2023). A good example of this in action is the Responsible AI Collective's AI Incident Database (AIID) (McGregor, 2021). The National Institute of Standards and Technology (NIST) also developed a hierarchical taxonomy that organizes AI risks into three main areas: technical design issues, social-technical issues, and guiding principles for building trust (NIST, 2023). Alongside these government efforts, organizations have also worked on ethical risk taxonomies. For example, the Alan Turing Institute developed a taxonomy to identify and organize different types of AI-related harm, helping people better understand ethical challenges in AI (Leslie, 2019). Microsoft created practical guidelines for its product teams to identify common types of harm. These guidelines help increase awareness and suggest actions for reducing harm during the development of AI technologies (Microsoft, 2023).

### 2.3. Gaps in existing studies and our approach to developing a comprehensive taxonomy

Despite ongoing efforts by academia, industry, and government to develop taxonomies and guidelines for AI incidents and risks, several gaps remain. First, existing taxonomies are derived from secondary sources, such as academic literature (e.g., Clarke & Whittlestone, 2022; Dafoe, 2024), industry policy documents (e.g., Leslie, 2019; Microsoft, 2023; Zeng et al., 2024), and governmental regulations (e.g., Gov.UK, 2024; National Institute of Standards and Technology (NIST), 2024; Zeng et al., 2024), rather than from real-world incident data. As AI technologies evolve rapidly, these taxonomies fail to capture how AI risks manifest in real-world contexts and neglect emerging issues that are not yet well-documented in these sources. **Second**, most existing taxonomies focus solely on summarizing types of risks (e.g., Lee et al., 2024; Shahriar et al., 2023; Weidinger et al., 2022), ignoring essential contextual factors such as root causes, responsible entities, and how incidents are disclosed. These taxonomies cannot capture the dynamics of AI incidents, such as how frequently certain incidents occur, and how likely certain entities are to cause incidents at a specific AI lifecycle stage. Understanding these dynamics is crucial for addressing the impacts of AI incidents and designing effective prevention (Turri & Dzombak, 2023). **Third**, the few taxonomies that considered contextual factors tend to be either narrowly focused on specific domains (e.g., Golpayegani et al., 2022, in public health) or rely on overly broad categorizations (Yampolskiy, 2016, on internal or external causes; Slattery et al., 2024, on intentional or unintentional incidents). These taxonomies fail to distinguish between incidents arising from diverse sources, such as AI algorithms, developers, and adopting entities (Slattery et al., 2024). Likewise, dividing incidents into pre- and post-deployment misses those that happen at other stages of AI development, like planning, preparing data, or building models (Ng et al., 2022; Shahriar et al., 2023). Other taxonomies group incidents by where they start (e.g., human-AI interaction or system design) or who is affected (e.g., users or the public), but these groups are often too simple to reflect the full complexity of AI incidents (Velázquez et al., 2024).

Building on prior AI incident taxonomies, our research presents a taxonomy empirically derived from a thematic analysis of $N = 202$ real-world AI privacy and ethical incident reports from 2023 and 2024. Our taxonomy extends beyond classifying incident types by integrating contributing contextual factors, such as the causes of incidents, responsible entities, sources of disclosure, and consequential impacts (see Figure 2). We also provide a more granular classification of incident types across specific AI lifecycle stages while expanding the scope of responsible entities to include not only AI algorithms and malicious human actors (Velázquez et al., 2024) but also organizational and governmental failures. We also classify the consequential impacts of AI incidents, ranging from single-user harm to societal disruption, to better capture their broader consequences. Our research offers a more nuanced understanding of AI incidents and provides actionable insights for policymakers, developers, and researchers to enhance AI governance, improve incident detection, and strengthen risk mitigation strategies.

## 3. Methodology

Our research develops a taxonomy of incident types and the contextual factors, grounded in empirical evidence from a thematic analysis of real-world AI privacy and ethical incidents. In this section, we present our methodology for data source selection, data collection, extraction, screening, analysis, and taxonomy construction.

### *3.1. Database selection and scope definitions*

We selected the AI, Algorithmic, and Automation Incident and Controversy Repository (AIAAIC)[1] as our primary data source. This decision was based on two reasons. First, the AIAAIC repository is widely recognized in AI incident literature as the largest and most up-to-date collection of crowd-sourced AI incidents (Lee et al., 2024; Raji et al., 2022; Turri & Dzombak, 2023). Second, after comparing with the AI Incident Database (AIID)[2] and the OECD AI Incidents Monitor (AIM),[3] we found that the AIAAIC repository includes the majority of AI-related incidents documented in these other sources. However, although the AIAAIC repository categorizes factors such as affected sectors and incident impacts (see Appendix B), its current taxonomy and the inconsistent terminologies from its open-source nature limit its ability to comprehensively and clearly document emerging AI risks and harms (Abercrombie et al., 2024). Our research addresses these limitations by providing a detailed analysis of AI incidents and their contextual factors. We include a definition of the AIAAIC repository metadata fields in Appendix B.

To understand the diverse contextual factors contributing to AI incidents, our research examined not only confirmed past privacy and ethical incidents involving AI systems but also issues currently under formal investigation by governmental authorities and those that have sparked public concern. Literature contributes AI to diverse technologies such as predictive algorithms, large language models, and robotics (Lee et al., 2024). For consistency in our research scope, we adopt the European Commission's AI definition that captures its diverse capabilities: *"a machine-based system that is designed to operate with varying levels of autonomy and that may exhibit adaptiveness after deployment,* *and that, for explicit or implicit objectives, infers, from the input it receives, how to generate outputs such as predictions, content, recommendations, or decisions that can influence physical or virtual environments*" (European Commission, 2024). We defined *AI privacy incidents* as those that create or amplify the 12 AI-specific privacy risks identified in prior research (Lee et al., 2024), rooted in a traditional privacy taxonomy (Solove, 2006) (see Appendix C). Similarly, we define *ethical AI incidents* as violations of the 11 AI ethical principles synthesized from global private and public sector guidelines (Jobin et al., 2019) (see Appendix D). These definitions guided our screening and analysis.

### 3.2. Data collection and extraction

We followed established methodologies for incident report analysis from prior research (e.g., Hadan et al., 2021; Lee et al., 2024; Turri & Dzombak, 2023) throughout our data collection, cleaning, screening, and thematic analysis. Figure 1 presents our research processes. On September 23, 2024, we downloaded metadata for 622 reports from the AIAAIC repository into a spreadsheet (see Appendix B). These reports documented incidents occurring between 2023 and 2024. From these, we selected 210 reports labeled by contributors as involving "privacy" ($n = 126$) and "ethical" ($n = 84$) issues into a dataset.

To ensure we captured the most up-to-date AI incidents at the time of our research, we continued monitoring the AIAAIC repository weekly until completing our data analysis on November 1, 2024. This process added $n = 6$ additional reports to our initial dataset, bringing the total to $N = 216$ reports, with $n = 210$ collected initially and $n = 6$ collected in subsequent weeks.

For each selected report, we manually retrieved its content from the AIAAIC repository website based on the report ID and URL in the metadata, saved the content as a PDF, and extracted the text using Adobe Acrobat PDF Reader. These 206 text files were then uploaded to Dovetail[4] for screening and analysis.

To validate the inclusivity of our selected reports, we randomly reviewed 20 unselected reports and found that while 12 matched with our research scope—indicating that relevant cases may exist among the unselected reports—these incidents presented similar patterns and themes with those in our selected dataset. This suggests that additional reports were unlikely to yield new insights and that our selected reports had reached data saturation. Our focus on explicitly labeled privacy and ethical incidents allowed for more targeted analysis of these specific concerns.

### 3.3. Data screening and analysis, and taxonomy construction

Two researchers with expertise in human-centered generative AI, privacy, ethical design, and VR research screened the reports on a case-by-case basis, based on the following *inclusion criteria*. Specifically, we included incidents that:

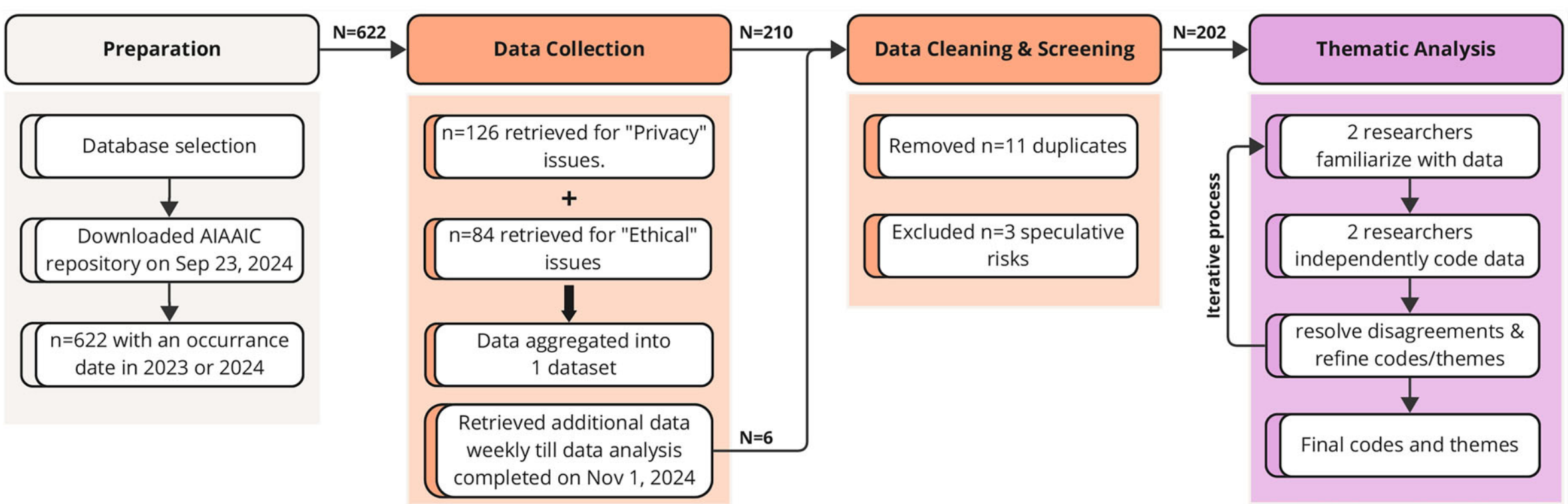

Figure 1. The flowchart shows our AIAAIC repository data collection process, from search keywords to data cleaning, screening, and final thematic analysis.

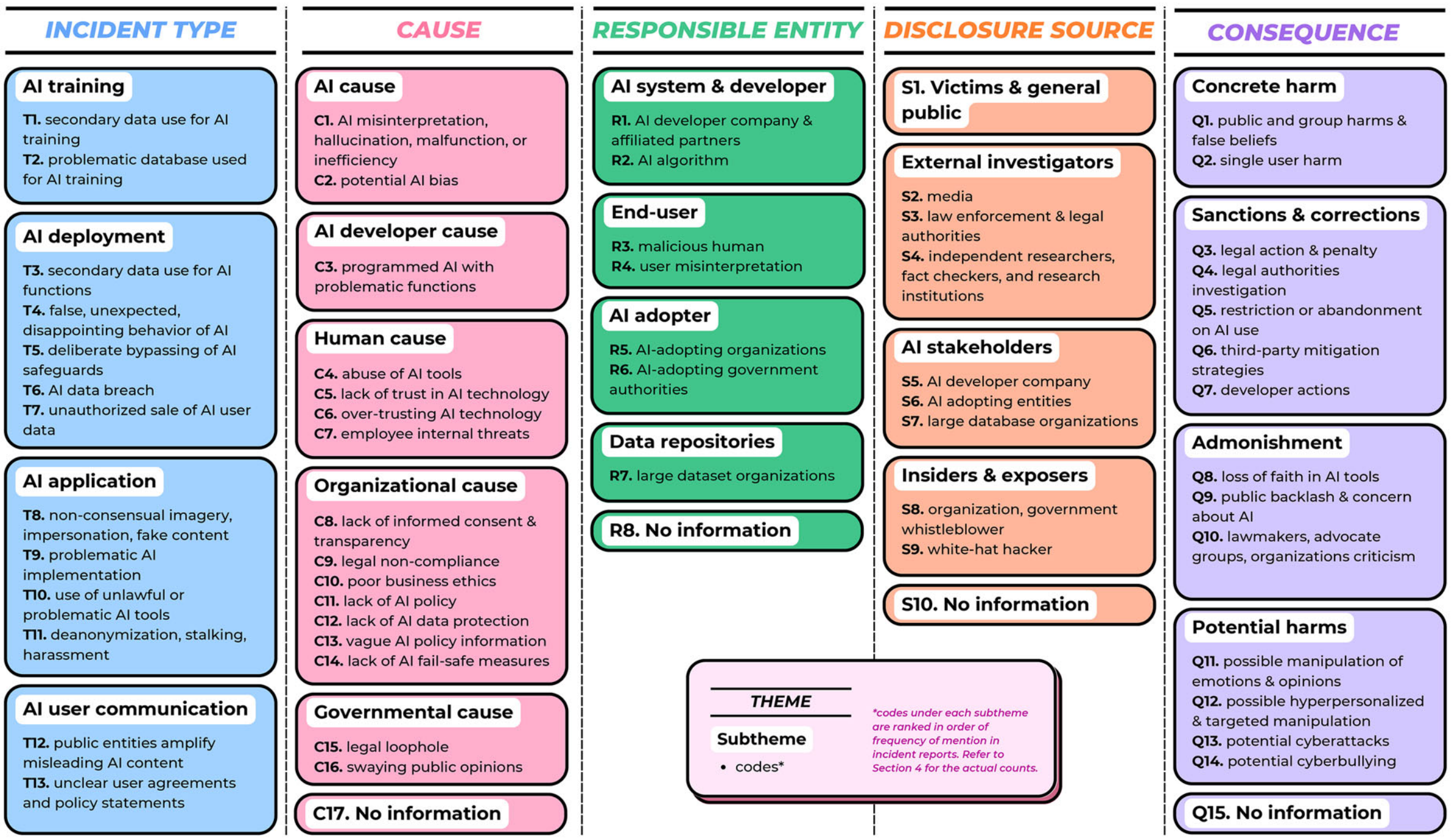

**Figure 2.** Our AI privacy and ethical incidents taxonomy (thematic analysis codebook). This figure presents our taxonomy developed through our thematic analysis, including the major themes, subthemes, and representative codes derived from incident data. The taxonomy contains 13 incident types spanning four stages of the AI development and deployment lifecycle (Ng et al., 2022; Shahriar et al., 2023). The taxonomy also includes six subthemes of 17 causes and five subthemes of eight responsible entities, covering not only end-users and AI algorithms but also developers, adopting organizations and government entities. It also includes five subthemes of sources of incident disclosure and five subthemes of 15 consequences, ranging from individual harms to collective and societal impacts.

1. involved systems that align with our adopted AI definition (see Section 3.1),
2. resulted in harms or posed risks aligned with our definitions of privacy and ethical AI incidents (see Section 3.1), and
3. led to harms or posed risks that were either under formal investigation by governmental authorities or had raised significant public concerns.

Our screening process excluded $n = 11$ duplicate reports with identical IDs and content. We further excluded $n = 3$ reports that only described the deployment or creation of AI systems without verified harms, risks, or public concerns, leaving $N = 202$ reports for thematic analysis. The thematic analysis was conducted by the two researchers who had already familiarized themselves with the data during screening. We used an integrated deductive-inductive approach that combines both deductive and inductive components during data analysis (Hadi Mogavi et al., 2022; Lyon et al., 2019). The deductive component involved constructing a set of themes based on recommended practices from prior AI research (Turri & Dzombak, 2023; Velázquez et al., 2024), cybersecurity incident analysis methodologies (Hadan et al., 2021; Serrano et al., 2019), and domain-specific AI risk taxonomies (Golpayegani et al., 2022; Hutiri et al., 2024). These themes guided our initial coding process. The inductive component complemented the deductive component by developing and refining codes, subthemes, and themes based on our data. This hybrid approach allowed us to start with a structured, literature-informed framework while deriving new insights from the specific context of our incident reports. Specifically, for each incident, we focused our analysis on the information relevant to five themes:

- The **type of incident** refers to the specific issue identified within an incident. In some cases, multiple issues may contribute to a single incident.

- The **cause** of the incident. In some cases, multiple causes may collaboratively contribute to an incident.
- The **responsible entity** in the incident. In some cases, multiple entities may be jointly responsible. For entities that could serve multiple roles (e.g., AI adopter, provider of large AI training datasets), we classified them based on their role and function in the AI incident.
- The **disclosure source** party that initially revealed the incident to the public. This could be the responsible entity itself, a third party (e.g., researchers, government agencies), or an individual affected by the incident. In some cases, multiple sources may be jointly involved in the incident disclosure.
- The **consequence** of the incident, or the harm, risk, concern, or actions the incident caused to users, organizations, or society.

Our thematic analysis began by randomly sampling $n = 33$ reports from the total of $N = 202$ selected reports. Two researchers independently read these reports line-by-line to assess the content in detail and created codes for the five themes and additional information that they observed from each specific report. Then, a meeting was held between the two researchers to discuss and resolve disagreements in the codes created and applied in these 33 reports. In the same meeting, they collaboratively reviewed the same 33 reports to ensure no insights were missed and refined and merged codes where needed. This process continued for an additional five weeks until all data were analyzed. Each week, both researchers independently coded a same portion of the remaining reports ($n = 34$ reports every week for four weeks and $n = 33$ reports for the last week). After each round of independent coding, the researchers met to review the same reports, ensure all insights in the data were captured, resolve any disagreements, and further refine and add to the codes. Finally, a review session was held to examine all codes, identify broader patterns across the dataset, and develop subthemes in addition to the five themes. After this session, the themes, subthemes, codes, and codebook were finalized.

Figure 2 summarizes our codebook, outlining its themes, subthemes, and codes derived from our thematic analysis. Appendix E provides descriptive statistics on the frequency of each code across the incident reports, while Appendix F presents the complete codebook, including definitions and exemplary incident reports for each code. Our analysis resulted in a codebook that synthesizes 13 incident types across four stages of the AI lifecycle (Ng et al., 2022; Shahriar et al., 2023). The codebook also includes five subthemes of causes and responsible entities, covering end-users, AI algorithms, developers, adopting organizations and government entities, with four subthemes of incident disclosure sources and consequences, ranging from individual harms to societal impacts. We discuss the role of these factors and their interconnections in Section 4. Our full dataset, our codebook, and interactive treemaps that illustrate the relationships between incident types and contextual factors are available at https://osf.io/swy5j/.

## 4. Taxonomy of AI privacy and ethical incidents

Our taxonomy synthesizes AI privacy and ethical incidents from 2023 to 2024, and their contributing contextual factors, such as causes, consequences, sources of disclosure, and responsible entities. Figure 2 summarizes these factors based on the 202 analyzed AI incident reports, with detailed descriptions and examples for each code provided in Appendix F. For clarity, quotes directly extracted from the incident reports are presented in italics and quotation marks (e.g., *"dark patterns"*, and percentages are calculated relative to the total number of incidents in our dataset ($N = 202$). We also avoided naming specific AI systems or entities involved in the incidents to maintain impartiality towards all organizations studied, as our focus is on understanding the AI incident rather than attributing blame.

### 4.1. Types of AI incidents

We classified the AI incidents in relation to four stages of the AI technology lifecycle (Ng et al., 2022): 1) training, 2) deployment, 3) application, and 4) user communication. This approach allowed us to better understand the specific stages where AI incidents are most likely to occur.

#### 4.1.1. AI incidents in training (n = 29, 14%)

AI incidents in training refer to issues that arise during the collection of training data and the development of AI models (Ng et al., 2022). Although AI development goes through various stages, such as concept design and requirement planning (Wickramasinghe et al., 2020; Ng et al., 2022), our analysis specifically focuses on training-related incidents because they represented the documented issues in our dataset. This focused scope allows for detailed examination of incidents directly tied to the training process. The first type involves the `secondary data use for AI training` (T1: $n = 28, 14\%$), where data originally collected for other purposes is repurposed for AI training (Solove, 2006). For example, several social media platforms have reused user profile data to train AI models without properly obtaining user consent (AIAAIC Repository, 2024a, 2024b). Among the 28 incidents, six (3%) were associated with *"dark patterns,"* such as designs that automatically opt users' data into AI training (AIAAIC Repository, 2024a, 2024c). These designs deceive users into sharing more data than they intended (Bösch et al., 2016).

The second type of incident involves `problematic database used for AI training` (T2: $n = 1, < 1\%$), which occurs when training data contains biased, inaccurate, or unrepresentative content. These contents can adversely impact the AI model's decision-making and outputs. Such incidents are relatively rare in our findings, with only one documented case where offensive, false, and biased content was incorporated into the training datasets of large language models developed by well-known AI companies (AIAAIC Repository, 2023b). Nonetheless, once unauthorized or biased data is included in AI training, it becomes difficult to remove. This problem is evident in five incidents (2%) in which developers' attempts to delete false information and correct the AI's problematic behavior were unsuccessful, as the issues reemerged after a period of time (e.g., AIAAIC Repository, 2023c, 2024d).

#### 4.1.2. AI incidents in deployment (n = 62, 31%)

AI incidents in deployment refer to issues that arise when AI systems are implemented in real-world environments (Ng et al., 2022). Through our analysis, we identified five types of incidents that commonly occur at this phase. The first type involves the `secondary data use for AI functions` (T3: $n = 32, 16\%$), where data is repurposed to power AI-driven features or services (Solove, 2006). For example, a regional police office was found to use real citizens' personal data to covertly test AI-powered analytics software, unaware that the software could bridge multiple databases and uncover information far beyond their intended scope of analysis (AIAAIC Repository, 2023d). In five cases (2%), we found that this type of incident triggered public complaints over the *"unjustified"* use of data and lawsuits against entities employing AI, as the data used to power AI capabilities was either accessed without user consent or deemed excessive for the specific services being offered (e.g., AIAAIC Repository, 2023e, 2023f).

The second type involves `AI false & unexpected & disappointing behavior` (T4: $n = 20, 10\%$), where an AI system behaves in ways that deviate from its intended functionality, produces incorrect or unreliable outputs, or fails to meet user expectations even when operating as designed. For example, one incident involved an AI chatbot falsely accusing a journalist of serious crimes due to its misinterpretation of the journalist's extensive career on court cases involving abuse and fraud (AIAAIC Repository, 2024d). Despite developers' attempts to address and rectify these false behaviors (as mentioned in Section 4.1.1), our analysis also revealed four cases (2%) where AI developers refused to address the false behavior and withheld details about the AI system's design, evaluation, and operation from government agencies (e.g., AIAAIC Repository, 2020b, 2024e). Such actions violate the accuracy principle of the General Data Protection Regulation (GDPR), which requires that users' personal data be maintained accurately (European Parliament and Council of the European Union, 2016).

The third type involves the `deliberate bypassing of AI safeguards` (T5: $n = 6, 3\%$), where individuals or groups exploit AI vulnerabilities, manipulate systems, or override built-in safety mechanisms and user agreements to achieve specific objectives. For example, one incident report described how the prompt injection technique was employed to manipulate an AI chatbot into disclosing users' precise location data, successfully bypassing its built-in safeguards intended to protect such information (AIAAIC Repository, 2023g).

In addition to the intentional bypassing of AI safeguards, we identified a fourth type of incident in AI deployment: `AI data breach` (T6: $n = 3, 1\%$). These occur when data used by or exposed to an AI system is compromised through unauthorized access or exploitation due to vulnerabilities in its algorithms, data storage, or operational infrastructure. An example is the breach of an AI-powered hiring chatbot, where hackers gained access to sensitive information about job applicants and the company itself (AIAAIC Repository, 2024f). Beyond data breaches, we also found incidents involving the `unauthorized sale of user data` (T7: $n = 3, 1\%$), where the organizations that maintain AI data purposefully monetize data from users. Examples of such data include user photos, video recordings, conversations stored on AI chatbots (AIAAIC Repository, 2023h), audio recordings and students' papers (AIAAIC Repository, 2024g), and browser data (AIAAIC Repository, 2023i). While AI was not the direct cause of these incidents, its extensive data collection capabilities likely made it an attractive target for hackers and monetization (Butterick, 2022; United States District Court Northern District Of California & Class Action Complaint, 2023).

#### 4.1.3. AI incidents in application (n = 124, 61%)

AI incidents in application refer to issues that arise when AI systems are used to perform specific tasks or provide functionalities directly to end users or within defined workflows. Our analysis revealed four types of AI incidents that commonly occur at this phase. The first type involves `non-consensual imagery, impersonation, fake content` (T8: $n = 78, 39\%$), which involves the use of AI tools, such as deepfake technology, to create hyper-realistic images, audio, or videos depicting individuals in fabricated scenarios. This type of incident is also the most prevalent, compared to other types of AI incidents (see Figure 3). Alarmingly, seven (3%) incidents involving both AI-based non-consensual imagery and harassment specifically targeted children. In these cases, students created and shared AI-generated nude images of their classmates and school staff (e.g., AIAAIC Repository, 2023j, 2024h). Beyond these personal harms, we identified 31 (15%) incidents where deepfakes posed risks to military, judicial, and political domains. For example, a deepfake photograph of an explosion near the Pentagon in Washington, D.C., caused a 0.26% drop in the U.S. stock market (AIAAIC Repository, 2023k). In another instance, a deepfake video of the Philippine president seemingly ordering a military attack on China put diplomatic relations at risk (AIAAIC Repository, 2024i). Even in everyday contexts, this type of incidents have increased public fears, as the technology can be exploited to target vulnerable populations and manipulate public opinions (e.g., AIAAIC Repository, 2023g, 2024j, 2024k). In seven (3%) reports, this type of incident further led to the `deanonymization, stalking, harassment` (T11: $n = 9, 4\%$), where AI tools were used to re-identify individuals from anonymized data by linking publicly shared images or datasets to a person's real identity. For example, a *"digital peeping Tom"* used a facial recognition platform to uncover the real identities of anonymous adult performers by uploading screenshots from films and tracking their online presence (AIAAIC Repository, 2023l).

The third type of AI incidents involves the `problematic AI implementation` (T9: $n = 30, 15\%$), which occurs when the application of AI tools leads to unintended harms, privacy and ethical concerns, or societal backlash. For example, an AI-powered tool designed to help users find previously viewed content in a major operating system raised privacy concerns because of its requirement to automatically capture screenshots of users' screens constantly (AIAAIC Repository, 2024l). Similarly, AI speed cameras, despite potential road safety benefits, faced criticism for being an invasion of driver privacy (AIAAIC Repository, 2024m). These incidents highlight the ongoing challenge of *"balancing privacy concerns with the potential benefits of AI"* innovation and functionality (AIAAIC Repository, 2023m, 2024n). While many of these technologies are developed with good intentions, they can inadvertently create risks, particularly for vulnerable populations. An example is AI-powered textbooks designed for classroom education, which raised concerns about *"potential negative impacts on children's development"* and an *"over-reliance"* on AI (AIAAIC Repository, 2024o).

The fourth type involves the `use of unlawful or problematic AI tools` (T10: $n = 17, 8\%$), where the applied AI systems were known to be controversial, unethical, or illegal. For example, a whistleblower from an autonomous vehicle company claimed that the company prioritized business growth over safety, continuing to market and deploy its self-driving technology despite being aware of its safety issues (AIAAIC Repository, 2023n). Similarly, several companies have received fines for AI-based employee surveillance (e.g., AIAAIC Repository, 2023o, 2024p), and government agencies have faced criticism for using privacy-invasive AI systems to monitor citizens, detect weapons, and track homeless populations (e.g., AIAAIC Repository, 2023p, 2024n). These incidents illustrate the use of problematic AI tools by both private companies and public government sectors, which are typically expected to prioritize ethical practices and citizen protection (see Figure 4).

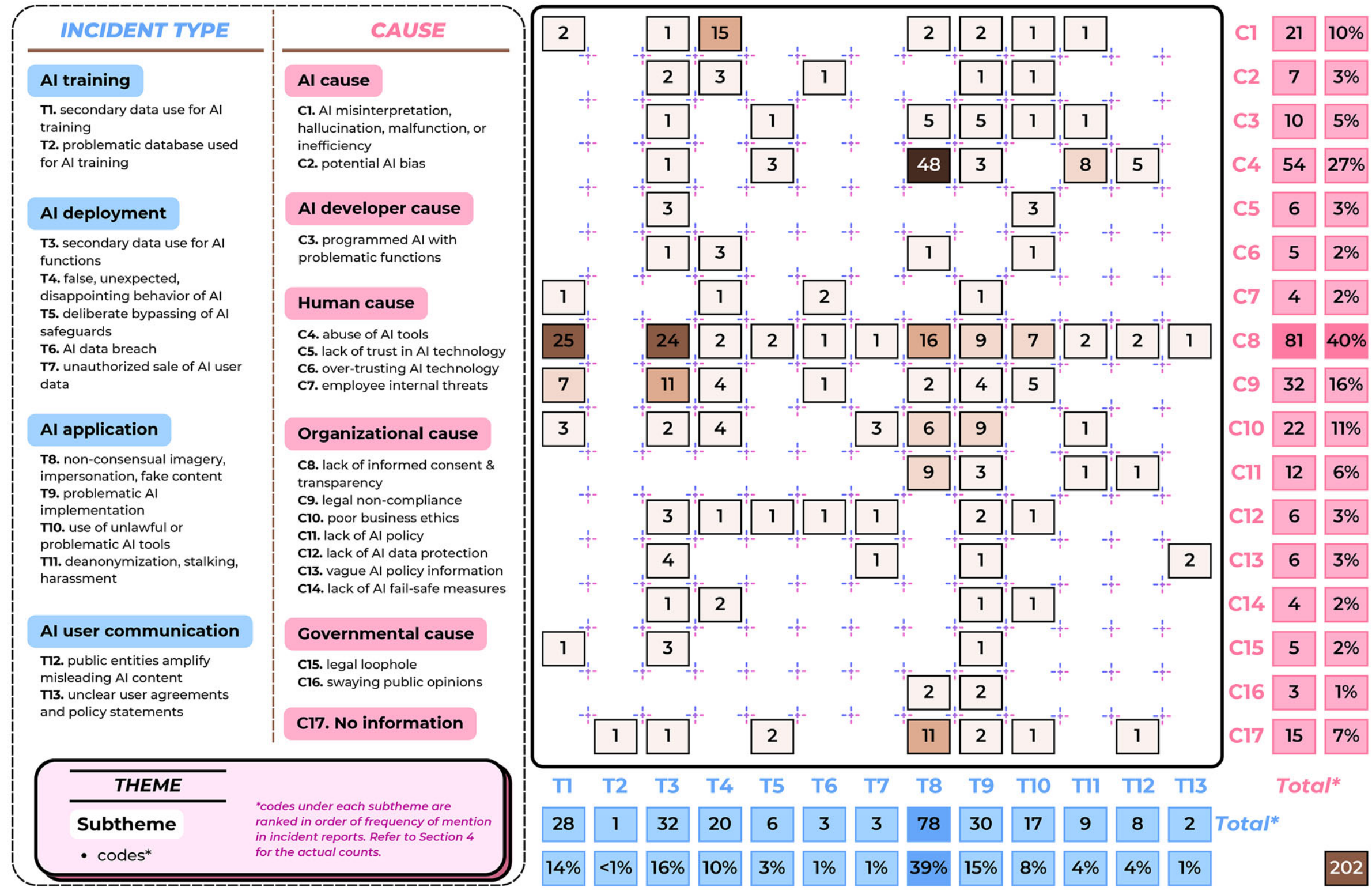

**Figure 3.** The heatmap that demonstrates the number of AI incidents and their corresponding causes. Each cell represents the number of incidents attributed to a specific cause. The horizontal axis lists the types of incidents (T), and the vertical axis indicates the causes (C). The legend on the left presents the name of each incident type and cause. Percentages are calculated based on $N = 202$. for example, we found 25 (12%) `secondary data use for AI training` incidents occurred because of `lack of informed consent & transparency`. Total counts $> 202$ and total percentage $> 100\%$ because incidents involving multiple issues or causes are coded with all relevant codes. *totals do not equal the sum of incidents because they represent the unique number of incidents within each theme/subtheme/code.*

#### *4.1.4. AI incidents in user communication ($n = 10$, 5%)*

AI incidents in user communication refer to issues that arise when the information, terms, or policies of AI systems are communicated to end users. Through our analysis, we identified two common types of incidents at this phase. The first type of incident involves `public entity amplify misleading AI content` (T12: $n = 8, 4\%$), where public figures inadvertently or deliberately share AI-generated articles, images, videos, or social media posts that misinform the public, distort facts, or spread propaganda. These incidents often blur the line between satire and facts, as the public's trust in these figures makes it less likely for them to verify the authenticity of the shared content, even when it includes deepfakes. For instance, a political figure was reported to have used AI to create a fake endorsement from a prominent celebrity for his presidential campaign, leading to potential impacts on public perception (AIAAIC Repository, 2024q).

The second type involves `unclear user agreements and policy statements` of AI systems (T13: $n = 2, 1\%$), where AI systems' terms of service, privacy policies, and consent mechanisms are complex, vague, or difficult for users to fully understand the implications of their interaction. For example, users of a major design software raised privacy and copyright concerns when service terms were updated with vague language about using customer content for AI development purposes (AIAAIC Repository, 2024r). In the other incident, a video conferencing platform's revised terms of service created confusion as it appeared to contradict the company's previous statement regarding AI training data usage (AIAAIC Repository, 2023q).

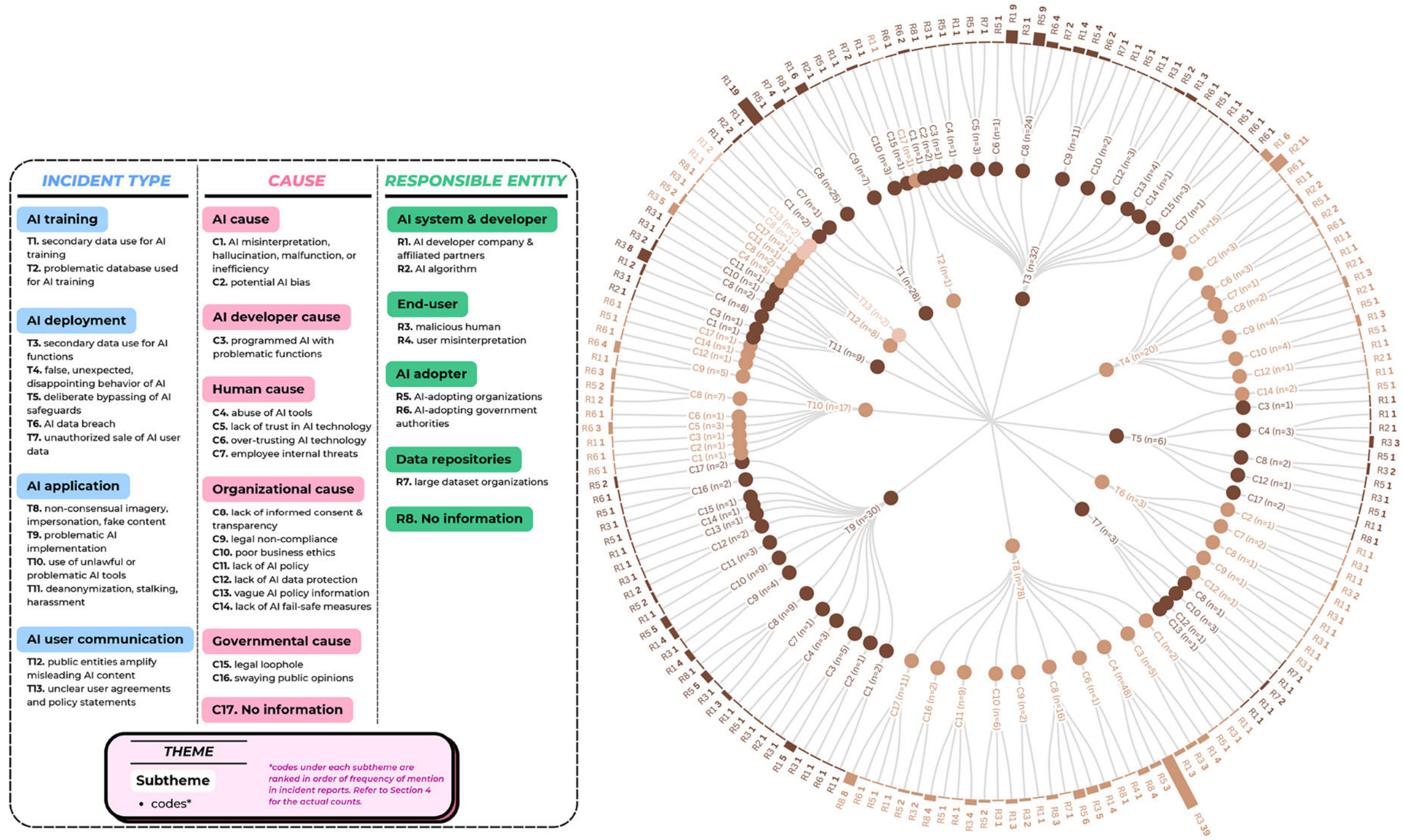

**Figure 4.** Radial treemap illustrating the relationship between AI incident types (T) and their corresponding causes (C) and responsible entities (R). Each bar at the outer edge represents the count of incidents attributed to a specific responsible entity within a particular cause. The legend on the left presents the name of each incident type, cause, and responsible entity. Longer bars indicate higher frequencies of incidents for a given cause and responsible entity. For example, we found 19 (9%) `secondary data use for AI training (T1)` incidents occurred because of `lack of informed consent & transparency (C8)` by `AI developer company & affiliated partners (R1)`. Incident reports involving multiple causes or types are coded with all relevant codes.

### 4.2. Common causes of AI incidents

To better understand the root causes of incidents and inform targeted interventions, we classified the causes of AI incidents in relation to the sociotechnical context of the AI technology, including: (1) AI technical causes, (2) AI developer causes, (3) human causes, (4) organizational causes, and (5) governmental causes (see Figure 2), each representing distinct but interrelated causes that contributed to AI privacy and ethical incidents.

#### 4.2.1. AI technical (n = 25, 12%) and developer causes (n = 10, 5%)

AI technical causes refer to causes that originate from the technical algorithm of AI systems. Through our analysis, we identified two causes that originated from the technical algorithm of AI systems. The first cause involves `AI misinterpretation, hallucination, malfunctions, inefficiency` (C1: $n = 21, 10\%$), where incidents occur because of errors and limitations in an AI system's performance, leading to incorrect, unreliable, or unintended outputs. Examples include AI systems generating false accusations and misinformation (e.g., AIAAIC Repository, 2024d, 2024e), misidentifying people as thieves or homeless (e.g., AIAAIC Repository, 2024n, 2024s), false positives from AI-powered detectors for fraud and plagiarism (e.g., AIAAIC Repository, 2020b, 2023r), and sensitive information leaks caused by AI hallucinations (e.g., AIAAIC Repository, 2023a). This cause is the most common cause of incidents involving `AI false & unexpected & disappointing behavior` by AI systems (see Figure 3).

Beyond performance errors, `potential AI bias` (C2: $n = 7, 3\%$) represents the second type of technical cause, where AI systems produce discriminatory outcomes or reinforce societal inequalities due to biased training data, or insufficient consideration of diverse perspectives during development. For example, facial recognition software has been shown to unfairly target women, people of color, or individuals from certain ethnic backgrounds (e.g., AIAAIC Repository, 2023s, 2024s). Predictive policing algorithms have disproportionately focused on minority communities (e.g., AIAAIC Repository, 2023t), and AI chatbots have been found to disseminate politically biased opinions (e.g., AIAAIC Repository, 2023u). These examples highlight that biased AI systems have the potential to amplify existing societal inequities.

In addition to incidents caused by AI algorithms, we identified 10 (5%) incidents caused by AI developers' decisions and practices. This occurs when AI systems are `programmed with problematic functions` (C3) that are considered unethical or privacy invasive. Examples include AI tools that enable employers to monitor workers' activities (e.g., AIAAIC Repository, 2024p), take frequent screenshots of users' screens (e.g., AIAAIC Repository, 2024l), allow and incentivize users for creating deepfakes of real people (e.g., AIAAIC Repository (2023v, 2024t), and denudification software (e.g., AIAAIC Repository, 2024w).

### 4.2.2. Human causes ($n = 69$, 34%)

Human causes refer to causes originating from the actions, decisions, or misunderstandings of humans interacting with AI systems. Through our analysis, we found four human causes of AI incidents. The first cause involves the `abuse of AI tools` (C4: $n = 54, 27\%$), where the incident occurred because of intentional use or exploitation of AI systems by individuals or groups to achieve unethical, illegal, or harmful objectives. For example, as mentioned in Section 4.1, malicious actors can exploit AI systems into disclosing users' data (AIAAIC Repository (2023g), predict users' demographic information based on their media posts (AIAAIC Repository (2023w), generate deepfakes for stalking and harassing (AIAAIC Repository, 2023j, 2024h, 2024v), and disinformation campaign (AIAAIC Repository, 2024w). This is the leading cause of incidents involving `nonconsensual imagery, impersonation, and fake content` (T8: $n = 48, 24\%$), and the second leading cause of all AI incidents ($n = 54, 27\%$; see Figure 3).

The second cause relates to the general public's `lack of trust in AI` (C5: $n = 6, 3\%$), where the incident occurred due to skepticism or fear surrounding AI technology. This is exemplified by opposition from over 56,000 parents, which delayed South Korea's plan to introduce AI-powered digital textbooks for personalized learning (AIAAIC Repository (2024o). Another example is the public criticism and privacy concerns regarding AI-powered speed cameras (AIAAIC Repository, 2024m).

The third cause involves `over-trusting AI` (C6: $n = 5, 2\%$), where the incident occurred because individuals or organizations rely blindly on the decisions, predictions, or recommendations made by AI systems without adequately verifying their accuracy. In one incident, a woman was misidentified by a facial recognition system as a banned shoplifter; despite presenting three forms of photo identification to prove her innocence, she was still accused of theft and asked to leave the store (AIAAIC Repository (2024s). In another incident, a child protection worker used AI to draft a critical custody case report without proper review, resulting in errors, including inappropriate language and sentence structures that were inconsistent with child protection protocols (AIAAIC Repository, 2023x).

Beyond the external human causes, we found four (2%) incidents caused by the malicious actions of employees within AI developer or adopter companies (`employee internal threats` [C7]). One example is that employees at a smart home security company were found to download and use private videos of female users in bedrooms and bathrooms because of their unrestricted access (AIAAIC Repository, 2023y). Another incident involves employees of an AI facial recognition company publicly posted biometric data, leading to the leak of over one million biometric records (AIAAIC Repository, 2024x). This is the most common cause of AI incidents involving data breach (T6: $n = 2$, $< 1\%$; see Figure 3).

#### 4.2.3. Organizational causes ($n = 117$, 58%)

Organizational causes originate from the policies, practices, or decision-making processes of organizations that develop, deploy, or oversee AI systems. Through our analysis, we identified seven organizational causes of AI incidents. The most prominent cause is the `lack of informed consent/transparency` (C8: $n = 81, 40\%$), where incidents occur because users are not adequately informed about how their data is collected, processed, or used by AI systems. This is also the leading cause of `secondary data use for AI functions` (T3: $n = 24, 12\%$) and `secondary data use for AI training` (T1: $n = 25, 12\%$; see Figure 3). Examples include using user data to train AI models without notifying or obtaining consent (AIAAIC Repository, 2023m, 2023q, 2024a, 2024b, 2024y), creating facial recognition databases from internet users without their consent (AIAAIC Repository, 2024z), using user data for AI-powered functionalities without notice (AIAAIC Repository, 2024i, 2024aa, 2024ab), deploying AI surveillance without disclosing data collection and use practices (AIAAIC Repository, 2023p, 2023y, 2024ac).

Similar to the lack of informed consent, we also identified incidents originated from `vague policy information` (C13: $n = 6, 3\%$), where existing policy statements were unclear and lack of details outlining how the AI system operate, use of user data, be governed, or address potential risks. This is the primary cause of incidents involving `unclear user agreements and policy statements` (T13: $n = 2$, $< 1\%$; see Figure 3). In these incidents, while efforts were made to establish policy statements, their lack of clarity often hindered users' understanding and reduced the effectiveness of communication. For example, updated terms of service allowing companies to use user data for AI improvement were found to be overly broad and vague (AIAAIC Repository, 2024r) or even contradictory to previous statements (AIAAIC Repository, 2023q). In another instance, despite a privacy notice stating that user conversations would never be shared with advertisers, behavioral data was still found to be shared and sold to advertisers (AIAAIC Repository, 2023h).

The third cause involves `legal non-compliance` (C9: $n = 32, 16\%$), where the incident occurred because the AI system was deployed in a manner that breached established legal standards. For example, a facial recognition AI company faced regulatory penalties in Europe for unauthorized collection of biometric data and non-compliance with users' data access requests (AIAAIC Repository, 2024z). As discussed in Section 4.1, several companies have faced fines or lawsuits for AI-based employee surveillance (e.g., AIAAIC Repository, 2023o, 2024p, 2024ad), and refusing to correct AI false behaviors while withholding details about the AI system's design, evaluation, and operation from government agencies (e.g., AIAAIC Repository, 2020b, 2024e). Commonly cited regulation standards in these incidents include, but are not limited to GDPR (European Parliament and Council of the European Union, 2016), the Texas Capture or Use of Biometric Identifier Act (CUBI) (CUBI, 2009), and the Illinois Biometric Information Privacy Act (BIPA) Illinois State Legislature (2008).

Beyond breaching legal standards, we identified a fourth type of organizational cause: major media platforms' `lack of AI policy` (C11: $n = 12, 6\%$). This includes the absence of effective guidelines, governance, or monitoring mechanisms to restrict and prevent the spread of harmful or fake AI-generated content. For instance, incidents such as the widespread deepfake video scams on social media (AIAAIC Repository, 2023z), deepfake videos of UK TV doctors circulating on major social media platforms (AIAAIC Repository, 2024ae), and a deepfake photograph of an explosion near the Pentagon in Washington, D.C. (AIAAIC Repository, 2023k). Additionally, one incident also revealed AI tools' lack of safeguards for children, such as displaying adult advertisements without implementing age restrictions (AIAAIC Repository, 2023f).

The fifth cause involves `lack of data protection` (C12: $n = 6, 3\%$), where incidents arise from failures to implement adequate security measures to safeguard AI systems and the sensitive data they process. Examples include weak password requirements on AI chatbots storing user photos, videos, and conversations (AIAAIC Repository, 2023h), incomplete data protection impact assessments required by law (AIAAIC Repository, 2021), and compromised security due to rushed product launches (AIAAIC Repository, 2024f).

In addition to insufficient data protection, we identified four (2%) incidents involving a `lack of AI fail-safe measures` (C14), where the absence of strategies and protocols to prevent or mitigate harmful outcomes from AI system malfunctions or unexpected behaviors contributed to the issue. This cause often serves as a secondary factor in incidents caused by `potential AI bias`. For example, a retailer failed to prevent false accusations of theft against innocent customers due to inadequate precautions when deploying facial recognition technology (AIAAIC Repository, 2023s). In another incident, a developer failed to establish the "*necessity and proportionality*" of the AI system, resulting in privacy violations (AIAAIC Repository, 2023t).

The final organizational cause pertains to `poor business ethics` (C10: $n = 22, 3\%$), where companies deliberately deploy untested, unsafe, or unethical AI systems to manipulate users, spread disinformation, or pursue profit at the expense of societal and environmental well-being. This is the most common cause of incidents involving the `unauthorized sale of user data` (T7: $n = 3, 1\%$; see Figure 3), including cases such as the sale of students' academic data (AIAAIC Repository, 2024g), user data collected from AI chatbot interactions (AIAAIC Repository, 2023i), and copyrighted data extracted through web browser (AIAAIC Repository, 2023i). Moreover, poor business ethics also frequently contribute to `problematic AI implementation` (T9: $n = 9, 4\%$; see Figure 3). For instance, companies have marketed and deployed self-driving technologies despite known safety issues (AIAAIC Repository, 2023n), incentivized users to create deepfakes of real people (AIAAIC Repository, 2023x), and ignored flaws in AI chatbots that posed harms to children (AIAAIC Repository, 2023g).

#### 4.2.4. Governmental causes ($n = 8$, 4%)

Governmental causes refer to incidents arising from the actions, policies, or regulatory gaps of government entities in the context of AI systems. Through our analysis, we identified two governmental causes. The first cause involves `legal loophole` (C15: $n = 5, 2\%$), where incidents occurred due to gaps in legal frameworks that allowed AI systems to be deployed without proper oversight, accountability, or safeguards. The absence of specific regulations in several jurisdictions has made it difficult to protect people from AI-generated deepfakes (AIAAIC Repository, 2024d, 2024af) or to regulate the unnecessary collection of biometric data by AI systems (AIAAIC Repository, 2023ab). Additionally, in regions lacking robust AI data protection laws, users often face challenges in opting out of having their data collected and used for AI training (AIAAIC Repository, 2024b).

Furthermore, we identified a second governmental cause in three incidents (1%), involving the deliberate use of AI systems by government authorities to `sway public opinions` (C16). These incidents occurred as a result of intentional efforts to influence, exploit, or alter individuals' thoughts, emotions, decisions, or behaviors. Notably, all such cases involved the political use of deepfakes, such as manipulating public opinion during elections (AIAAIC Repository, 2023ac, 2023ad) and influencing perceptions related to military activities (AIAAIC Repository, 2023ae).

Lastly, we also found 15 (7%) incidents where `no information` (C17) of their cause was provided, with only the description of the incidents being documented (see Figure 3).

### 4.3. Responsible entities and roles

In this section, we summarize the roles and responsibilities of the entities accountable for AI incidents based on our analysis results (see Figure 4). Our findings revealed seven responsible entities for AI incidents in four groups: (1) AI systems and developers, (2) end-users, (3) AI adopting organizations and government entities, and (4) data repositories.

The first group of responsible entities are AI systems and developers ($n = 77, 38\%$), which includes `AI systems' algorithms` (R2: $n = 14, 7\%$) that process input data to generate outputs, and `AI developer companies and affiliated partners` (R1: $n = 67, 33\%$) who design, develop, deploy, or maintain AI systems. As shown in Figure 4, `AI systems' algorithms` (R2) are frequently responsible for incidents involving `AI false & unexpected & disappointing behavior` by causing `AI misinterpretation, hallucination, malfunctions, inefficiency` (T4: $n = 11, 5\%$). On the other hand, `AI developer company and affiliated partners` (R1) represent the largest responsible entity among all we analyzed AI incidents ($n = 67, 33\%$) and is primarily responsible for incidents related to `secondary data use for AI training` (T1: $n = 21, 12\%$), `secondary data use for AI functions` (T3:

$n = 12, 6\%$), and unclear user agreement and policy statement (T13: $n = 2, < 1\%$). These issues are largely driven by the entity's lack of informed consent & transparency (C8: $n = 38, 21\%$), legal non-compliance (C9: $n = 18, 9\%$), and vague policy information (C13: $n = 5, 2\%$). Furthermore, we also identified the AI developer company and affiliated partners (R1) as secondary contributors to incidents originating from employee internal threats (C7: $n = 3, 1\%$). In these cases, failures in managing employee access to AI system data allowed them to become threats (e.g., AIAAIC Repository, 2023y). This entity further causes incidents involving problematic AI implementation (T9: $n = 12, 6\%$) and nonconsensual imagery, impersonation, fake content (T8: $n = 11, 5\%$) as they program AI with problematic functions (T3: $n = 8, 4\%$), such as those incentivize users to generate deepfakes (AIAAIC Repository, 2023v, 2024t). While these incidents were directly caused by end-users, the developers held partial responsibility for providing tools with inherently problematic functionalities. Lastly, this entity was also associated with an incident ($< 1\%$) involving problematic database used for AI training (T2), though the specific cause was not detailed in the report (AIAAIC Repository, 2023b).

The second group of responsible entities is end-users ($n = 52, 26\%$), which includes malicious human (R3: $n = 51, 25\%$) who intentionally exploit or misuse AI systems, as well as users who falsely interpreted the AI system's outcomes (user misinterpretation [R4]: $n = 1, < 1\%$). Our analysis revealed that malicious human (R3) is primarily responsible for incidents involving non-consensual imagery, impersonation, fake content (T8: $n = 44, 22\%$) and de-anonymization, stalking, harassment (T11: $n = 8, 4\%$). These incidents often originate from their abuse of AI tools (C4: $n = 43, 21\%$) and their exploitation of media platforms' lack of AI policy (C11: $n = 4, 2\%$) to conduct their malicious activities without restrictions. In cases of AI data breach (T6: $n = 3, 1\%$), this entity holds primary responsibility both for internal threats posed by employees within AI data management organizations (employee internal threats [C7]: $n = 2, 1\%$), and external hacking attempts the deliberate bypassing of AI safeguards (T5: $n = 3, 1\%$) that exploits AI systems' weakness. Public figures, such as celebrities, business leaders, media personalities, and political entities, also contribute to incidents involving public entity amplify misleading content (T12: $n = 5, 2\%$), as part of their deliberate efforts to swaying public opinions (C16: $n = 2, 1\%$). Moreover, user misinterpretation was responsible for the incident (R4: $n = 1, < 1\%$) involving non-consensual imagery, impersonation, fake content (T8) because of their over-trusting AI (C6) generated content as true, even when it was explicitly labeled as satire (AIAAIC Repository, 2023af).

The third group of responsible entities is AI adopting organizations and government entities ($n = 58, 29\%$), consisting of AI-adopting organizations (R5) that integrate AI systems into their operations as tools or services, or allow for the use of AI by its users, and AI-adopting government authorities (R6) that implement AI systems to assist with decision-making, managing services, or supporting public programs. We found that AI-adopting organization (R5) is primarily responsible for incidents involving problematic AI implementation (T9: $n = 13, 6\%$), often driven by their poor business ethics (C10: $n = 9, 4\%$), exploitation of legal loophole (C15: $n = 1, < 1\%$), and deliberate attempts to swaying public opinions (C16: $n = 1, < 1\%$). The impacts of these incidents are further exacerbated by the organizations' lack of AI fail-safe measures (C14: $n = 1, < 1\%$). On the other hand, AI-adopting government authorities (R6) were identified as the primary responsible entities in incidents involving use of unlawful/problematic AI tools (T10: $n = 12, 6\%$). These tools were often perceived as problematic by the general public (lack of trust in AI: C5: $n = 3, 1\%$) and the situation was further exacerbated by the potential AI bias (C2: $n = 1, < 1\%$) in the generated outputs (e.g., AIAAIC Repository, 2024ag).

The last responsible entity is large dataset organization (R7), such as data repositories, social media platforms, and online forums, which aggregate and manage extensive datasets used for training AI systems. We found that this entity was primarily responsible for incidents involving the unauthorized sale of user data for AI-related purposes (T7: $n = 2, 1\%$).

### 4.4. Source of disclosure

To understand how AI incidents come to public attention, we classified disclosure sources into four groups: (1) victims and the general public, (2) external investigators and authorities, (3) AI development and application stakeholders, and (4) insiders and exposers (Figure 5). In addition, we identified 14 (7%) incidents where the disclosure source was not specified in the report. In this section, we provide a brief overview of each disclosure source with examples identified from the incident reports.

The first group of AI incident disclosure sources consists of `victims and the general public` (S1: $n = 76, 38\%$), which include direct victims or individuals indirectly impacted by AI incidents (e.g., AIAAIC Repository, 2024d, 2024o, 2024r, 2024s), as well as third-party witnesses, such as users of digital platforms where problematic AI-generated content was shared (e.g., AIAAIC Repository, 2023z, 2024c) and users of AI systems who observed its leak of others' information (e.g., AIAAIC Repository, 2023a). Among the four identified disclosure source groups, this group accounted for the largest share of AI incident disclosures, constituting 38% of all reported cases. It also served as the primary disclosure source for five types of incidents, including those involving `problematic AI implementation` (T9: $n = 37, 18\%$), `secondary data use for AI training` (T2: $n = 12, 6\%$) and `nonconsensual imagery, impersonation, fake content` (T8: $n = 12, 6\%$), especially those associated with `AI developer & affiliated partners` (R1: $n = 22, 11\%$), `malicious human` (R3: $n = 19, 9\%$), `AI-adopting organizations` (R5: $n = 19, 9\%$) and `AI algorithm` (R2: $n = 5, 2\%$).

The second group of AI incident disclosure sources is external investigators and authorities ($n = 108, 53\%$). This group includes `mass media` (S2: $n = 49, 24\%$) such as journalists (e.g., AIAAIC Repository, 2023h) and news broadcasters and websites (e.g., AIAAIC Repository, 2023aa, 2024af); `law enforcement & legal authorities` (S3: $n = 46, 23\%$), including police and crime investigators (e.g., AIAAIC Repository, 2023t, 2024v), judicial authorities (e.g., AIAAIC Repository, 2023o, 2024n), regulatory bodies like data protection authorities and privacy commissioners (e.g., AIAAIC Repository, 2023m, 2023ag), and general government bodies like presidential communication office (e.g., AIAAIC Repository, 2024i); and `researchers & fact checkers` (S4: $n = 22, 11\%$), including research groups and institutions (e.g., AIAAIC Repository, 2023ad, 2024j), independent experts (e.g., AIAAIC Repository, 2023af), and non-governmental organizations dedicated to AI incident detection (e.g., AIAAIC Repository, 2020, 2023ah). These entities are the primary disclosure source for AI incidents involving `secondary data use for AI functions` (T3: totaled $n = 21, 10\%$), `use of unlawful/problematic AI tools` (T10: totaled $n = 12, 6\%$), and `problematic database used for AI training` (T2: totaled $n = 2, 1\%$), especially those responsible by `AI-adopting government authorities` (R6: totaled $n = 13, 6\%$), `large database organization` (R7: totaled $n = 7, 3\%$), and `user misinterpretation` ($n = 1, < 1\%$). We note that when individuals from these sectors are direct victims or affected parties in an incident, we categorized them in the first group of disclosure sources based on their role in the incident rather than their profession.

The third group of AI incident disclosure sources consist of stakeholders closely involved in AI development and application ($n = 13, 6\%$). These stakeholders include `AI developer companies` (S5: $n = 6, 3\%$) (e.g., AIAAIC Repository, 2024ah), `AI adopting entities` (S6: $n = 4, 2\%$) such as private commercial organizations (e.g., AIAAIC Repository, 2024x, 2024ai) and educational institutions (e.g., AIAAIC Repository, 2023r), and `large database organizations` (S7: $n = 3, 1\%$) like social media platforms and online forums where content was scraped without authorization for AI model training (e.g., AIAAIC Repository, 2023ai, 2024y).

The fourth group of AI incident disclosure source consists of insiders and exposers ($n = 5, 2\%$), which include `whistleblower` (S8: $n = 3, 1\%$) from organizations responsible for implementing problematic AI systems (e.g., AIAAIC Repository, 2023n), and `white hat hacker` (S9: $n = 2, 1\%$) who proactively evaluate AI system vulnerabilities before they can be maliciously exploited (e.g., AIAAIC Repository, 2023g, 2024f).

### 4.5. Consequences of AI incidents

To evaluate the scale and scope of harm caused by AI incidents, we categorized their consequential impacts into four subthemes: (1) incidents that caused concrete harm, (2) those that led to sanctions or corrections for AI systems or applications, (3) incidents that elicited formal admonishments, and (4) those posing potential risks for future harm. Additionally, we identified 18 (9%) that reports provided only a description of the incident, with `no information (Q15)` mentioning the consequence or impacts.

#### 4.5.1. Concrete harms ($n = 90$, 45%)

As shown in Figure 2, AI incidents result in two types of concrete harms: societal or collective damage (`public and group harms, false beliefs` [Q1]: $n = 44, 22\%$) and negative impacts on individuals (`single user harm` [Q2]: $n = 56, 28\%$). For instance, victims of AI data breaches and unauthorized data use have faced privacy loss (AIAAIC Repository, 2023x, 2023y), and those targeted by AI-powered deanonymization have lost online anonymity (AIAAIC Repository, 2023l). Deepfake victims often suffer from reputational damage (AIAAIC Repository, 2024w, 2023ah), financial loss (AIAAIC Repository, 2024j), and severe emotional distress, including humiliation, fear, and long-term psychological trauma (AIAAIC Repository, 2023j, 2024k, 2024af). Falsified AI outputs have triggered public panic over fabricated military attacks (AIAAIC Repository, 2024i) and created an illusion of false public opinion trends (AIAAIC Repository, 2023z). Misidentification by AI facial recognition systems has led to individuals' abduction, interrogation, physical abuse, and false accusations, depriving victims of their freedom (AIAAIC Repository, 2023s, 2024aj). Additionally, AI malfunctions have resulted in denial of seniors' pensions and welfare benefits (AIAAIC Repository, 2020b). People subjected to AI surveillance experienced intrusions on personal privacy and autonomy (AIAAIC Repository, 2023o), and employees under AI monitoring reported accelerated stress and workload intensity (AIAAIC Repository, 2024p).

#### 4.5.2. Sanctions or corrections ($n = 74$, 37%)

Through our analysis, we identified five punitive actions or corrections that resulted from AI incidents. The first is `legal action or penalty` (Q3: $n = 42, 21\%$), which includes lawsuits (e.g., AIAAIC Repository, 2023e, 2023w), fines (e.g., AIAAIC Repository, 2021, 2024z, 2024ad), criminal arrests and expulsions (e.g., AIAAIC Repository, 2024h, 2024x), and formal objections (e.g., AIAAIC Repository, 2023g, 2023m) imposed on individuals, organizations, or government authorities that were responsible for the incidents. Additionally, in three incidents (1%), we observed `legal authorities investigations` (Q4), where regulatory bodies, law enforcement, or other legal entities announced their upcoming official investigation into the AI system or its developers and operators (e.g., AIAAIC Repository, 2023n, 2023ag, 2023aj), even though clear legal violations were not apparent at the time of the incident report.

The third type of corrective actions is `restriction or abandonment on AI use` (Q5: $n = 17, 8\%$), which includes discontinuing, restricting, or prohibiting AI systems or specific functionalities. This can occur voluntarily by developers or adopting organizations and government entities after identifying issues (e.g., AIAAIC Repository, 2024ah), due to public complaints and boycotts (e.g., AIAAIC Repository, 2023r, 2024c, 2024o), or as a result of legal sanctions (e.g., AIAAIC Repository, 2023h, 2023s, 2024z).

Beyond complete prohibitions, we identified 20 (`developer actions` [Q7]: 10%) incidents where developers implemented corrective or protective measures in response to AI incidents instead of fully banning the systems. These measures included automatic classifiers to detect and block problematic outputs (e.g., AIAAIC Repository, 2024ah), privacy controls allowing users greater control over their data (e.g., AIAAIC Repository, 2023ak), security enhancements to address existing bugs and prevent future cyberattacks (e.g., AIAAIC Repository, 2023al, 2024f), updated privacy policies to clarify AI-related data practices (e.g., AIAAIC Repository, 2024l), data minimization efforts to delete outdated data (e.g., AIAAIC Repository, 2023y), and improved internal data management procedures to mitigate insider threats (e.g., AIAAIC Repository, 2023y).

When developers or adopting organizations and government entities fail to act, or in cases where AI incidents are initiated by external malicious actors, third-party entities—such as independent organizations, advocacy groups, or individuals—may intervene to address or mitigate the harms (`third-party mitigation strategies` [Q6]: $n = 4, 2\%$). For example, a third-party website was launched to support victims of deepfake pornography by providing resources and assistance (AIAAIC Repository, 2023j). In another incident, social media platforms removed problematic AI content and blocked associated accounts (AIAAIC Repository, 2023aa).

#### 4.5.3. Admonishment (n = 111, 55%)

Our analysis identified three types of admonishment resulting from AI incidents. The most common is `public user backlash or concern about AI` (Q9: $n = 91, 45\%$), including user criticism, protests, boycotts, or reduced usage of AI products and services. These reactions often occur due to public concerns about AI misuse in harassment, politics, or military actions (e.g., AIAAIC Repository, 2024i, 2024v, 2024w). Users also expressed frustration at the lack of AI governance, citing insufficient laws and regulations for AI incidents (e.g., AIAAIC Repository, 2023b, 2023ab), as well as the failure of major social media platforms to effectively detect, moderate, and control their spread (e.g., AIAAIC Repository, 2023z, 2024ae). Concerns also raised around AI systems' lack of transparency, inadequate security measures, privacy violations (e.g., AIAAIC Repository, 2024a, 2024r, 2024ah), and the potential to exacerbate racial and ethnic biases (e.g., AIAAIC Repository, 2023t, 2024s) were similarly prevalent. Parents also worried about risks associated with children's excessive use of AI, such as its potential negative impact on personal development, overall well-being (e.g., AIAAIC Repository, 2024o), and physical safety (e.g., AIAAIC Repository, 2023n, 2023o). Criticism also emerged around copyright infringement issues, as well as the adverse effects of AI on the livelihoods of artists, writers, actors, and musicians (e.g., AIAAIC Repository, 2023b, 2024k).

In addition to public backlash, we also identified `lawmakers, advocates groups, organizations criticism` (Q10: $n = 29, 14\%$) that reflected condemnation and proposed solutions from government officials, industry watchdogs, and independent groups. These entities pointed out the need for stronger and more effective AI governance, including measures to combat fake news, misinformation, and disinformation (e.g., AIAAIC Repository, 2024i), ensure data privacy and ownership rights (e.g., AIAAIC Repository, 2023i, 2024b), and address unethical practices like *"dark patterns"* in user experience design that make opting out of data collection difficult (e.g., AIAAIC Repository, 2024c). They also called for validation of claims made by AI developers and adopting organizations and government entities regarding the accuracy and safety of their systems (e.g., AIAAIC Repository, 2023n) and cautioned against the potential of synthetic media to confuse children and teenagers by blurring the line between fiction and reality (e.g., AIAAIC Repository, 2023aj). There was also emphasis on the importance to prevent the criminalization of AI (e.g., AIAAIC Repository, 2023ab, 2024n) and safeguard election integrity from AI-driven interference (e.g., AIAAIC Repository, 2024w).

The third type of admonishment is an erosion of trust (`loss of faith in AI tools` [Q8]: $n = 4, 2\%$), evidenced by a decline in public trust and confidence toward technology companies as a whole (AIAAIC Repository, 2024r) and increased concerns about the reliability of AI systems and the companies' ability or willingness to effectively monitor their platforms (AIAAIC Repository, 2023r, 2024e).

#### 4.5.4. Potential harms (n = 10, 5%)

We identified four types of potential future harms resulting from AI incidents. The first involves `possible emotional, opinions manipulation` (Q11: $n = 5, 2\%$), where AI systems create opportunities to influence people's emotions, beliefs, or decisions. For instance, AI companion apps may foster deep emotional connections with lonely individuals, encouraging them to share intimate details and leading to dependency or unrealistic expectations (AIAAIC Repository, 2023h). Similarly, deepfakes exploit the trust people place in experts, spreading misinformation more effectively by mimicking authoritative figures (e.g., AIAAIC Repository, 2024ae).

The second type is `possible hyper-personalized/targeted manipulation` (Q12: $n = 2, 1\%$), where AI systems deliver highly tailored content to influence users' decisions or behaviors.

For example, AI-powered vending machines recommending products based on user data may unfairly prioritize certain products or inflate their prices for profits (e.g., AIAAIC Repository, 2024aa, 2024ai).

The third type of potential harm is AI systems' potential to enable `cyberbully` (Q14: $n = 3, 1\%$) as AI allows for the automatic creation and rapid dissemination of harmful, targeted, or abusive content, amplifying the impact of cyberbully (e.g., AIAAIC Repository, 2024v, 2024ac). Lastly, we also found one ($< 1\%$) incident where researchers demonstrated how six commercial AI tools could be manipulated to facilitate `potential cyberattacks` (Q13), such as generating code for system breaches, data theft, database tampering, or denial-of-service attacks (AIAAIC Repository, 2023al).

## 5. Discussion

Our research provides a comprehensive taxonomy of AI privacy and ethical incidents, which is empirically derived from a thematic analysis of 202 real-world reports between 2023 and 2024. Our taxonomy classifies AI incident types and their contextual factors, including incident causes, responsible entities, disclosure sources, and consequential impacts. These contextual factors are important for understanding the reasons behind these incidents, identifying accountability, and assessing their effects on individuals and society, which further offers actionable insights for researchers, developers, and policymakers to better prevent and mitigate risks proactively and responsibly.

In this section, we discuss the implications of our findings in relation to existing literature, highlight the practical applications of our taxonomy, and illustrate its potential to inform AI governance, risk mitigation strategies, and ethical design practices. We also discuss the limitations of our study and propose directions for future research to further advance the understanding of AI risks and harms.

### *5.1. Comparing with existing AI privacy and ethical incidents taxonomies*

To demonstrate the inclusiveness and novelty of our taxonomy, in this section, we compare our taxonomy with AI privacy and ethical risks and incidents hypothesized or identified in the literature.

Our research extends prior literature by categorizing AI incidents across the four stages of the AI lifecycle (Ng et al., 2022), which offers a more detailed and actionable perspective on their occurrence. We also empirically synthesized the relationships between incident types, contextual factors, and their occurrence frequency in real-world contexts. This approach supports more targeted prevention strategies compared to studies that either did not address AI incidents within the lifecycle (e.g., Zeng et al., 2024), only distinguished pre- and post-deployment incidents (e.g., Slattery et al., 2024; Yampolskiy, 2016), or ignored the contextual factors (e.g., Shelby et al., 2023; Zeng et al., 2024).

Similar to previous studies (e.g., Slattery et al., 2024; Weidinger et al., 2022), we identified privacy compromises such as secondary data use, unauthorized sale of user data, AI-generated false behaviors, and the use of AI for deanonymization (Slattery et al., 2024; Zeng et al., 2024). However, unlike prior research that attributed privacy compromises primarily to AI systems (Slattery et al., 2024), our analysis reveals that AI developer companies and adopting entities held major responsibility for these incidents. In addition, we also observed real-world incidents aligned with previous literature on AI system security vulnerabilities and attacks (Lee et al., 2024; Slattery et al., 2024; Yampolskiy, 2016), such as external malicious actors bypassing AI safeguards, using AI for unethical and harmful purposes (e.g., disinformation, harassment, scams), conducting data security breaches, internal threats from AI developer and adopter employees, and the inadequacy of protections within the AI systems.

In addition, our research highlights the role of developers and adopting organizations and government entities in enabling or failing to mitigate AI incidents. While previous studies attributed misinformation to AI's false behavior (Slattery et al., 2024), we found that such behavior often originates from developers' decisions to use problematic training data, and exacerbated by their reluctance or inability to correct the issue. In fact, human abuse of AI systems, intentional use of deepfakes by adopting entities, and deliberate actions by public figures to sway public opinions were the leading causes of misinformation among our analyzed incidents. We also found incidents involving unsafe AI systems due to poor business ethics and media platforms' lack of policies on AI-generated content, reflecting the lack of oversight discussed in the literature (Cummings, 2024; Yampolskiy, 2016). Additionally, we observed 10 (5%) cases of inadequate AI design (Cummings, 2024; Yampolskiy, 2016), where AI systems were programmed with inherently problematic functions.

At the same time, our analysis also uncovered novel incidents and contextual factors that were outside the scope of prior literature or were insufficiently explored. For example, while prior work has primarily associated secondary data use with AI training (Lee et al., 2024), our findings revealed that secondary data use for facilitating AI functions (not training) is an equally large incident type ($n = 28, 14\%$ and $n = 32, 16\%$, respectively) among the incidents we analyzed. We believe it is necessary to separate them as these incidents happen in different stages of the AI lifecycle, and therefore, a clear separation can facilitate their better prevention.

While prior literature has emphasized socioeconomic harms from AI incidents, such as job loss, threats to autonomy, and diminished well-being (Abercrombie et al., 2024; Shelby et al., 2023; Slattery et al., 2024), our analysis shows a notable shift in impact patterns, with only 45% of incidents involving concrete harms. The majority of documented incidents resulted in reputational and operational consequences for AI developers and adopting organizations and government entities, including public backlash, legal penalties, and system abandonment. This shift suggests an evolution in how AI incidents manifest and are addressed, potentially driven by increased public scrutiny, stronger regulatory frameworks, and growing organizational accountability for AI systems. This finding contributes to our understanding of how AI incident impacts are changing as the technology matures and becomes more integrated into organizational operations.

As our analysis focused on the happened AI incidents, certain forward-thinking incidents and hypothetical incidents proposed in prior studies did not present in our analysis. For example, while prior studies have highlighted incidents arising from overreliance on AI systems (Slattery et al., 2024), such cases were relatively rare ($n < 10$) in our analysis. Another example is the hypothesis of AI incidents from intelligence overflow, when AI systems are so much ahead of us and no longer need command or communicate with human but operate and evolve in a way that attempt to intentionally harm human and society based on their understanding of protecting humanity (Slattery et al., 2024; Yampolskiy, 2016).

We also note that certain types of AI incidents and contextual factors, while identified in the literature, were absent or underrepresented in our analysis due to low disclosure rates from AI developers and adopting organizations and government entities. For example, while prior work discusses detailed AI privacy incidents from database management and security (e.g., model inversion attacks, derivation attacks) (Shahriar et al., 2023), these were not adequately captured in our taxonomy. Such incidents often involve internal processes that are only known to the procedures AI developer and adopting organizations and government entities. Similarly, incidents related to AI maintenance and testing (Cummings, 2024) were not fully captured in our analysis. While our taxonomy captures AI incidents and responsible entities from publicly available reports, we acknowledge an inherent limitation in accessing internal organizational processes. Our analysis can identify and categorize developer and adopter involvement in incidents, but cannot fully untangle the internal decision-making processes, management practices, or organizational factors that contributed to these incidents. This limitation reflects the broader challenge of studying AI development practices, where internal processes often remain opaque to external researchers. Rather than weakening our findings, this limitation highlights the need for future research combining public incident analysis with organizational case studies to better understand the full context of AI incidents.

### 5.2. Summary of insights from our analysis

Our analysis revealed that most reported AI issues occurred post-deployment, specifically during the deployment ($n = 62, 31\%$) and application stages ($n = 124, 61\%$). While this finding could partially reflect the AIAAIC repository's broader focus beyond developer-reported incidents, the systematic lack of pre-deployment incident reporting by AI developers and adopters represents a significant gap in understanding early-stage AI risks. This pattern mirrors similar challenges observed in other technological domains (Serrano et al., 2019) where pre-deployment incident reporting has been crucial for preventing downstream harms (Schultz & Shumway, 2001). The limited visibility into pre-deployment incidents suggests a need for both improved reporting mechanisms and potential regulatory frameworks to encourage transparency during AI development stages. As outlined in Section 4.4 the majority of incidents were brought to public attention by external sources such as victims, the general public, independent investigators, or regulatory authorities, rather than by those directly involved in AI development or management. Entities closely involved in AI systems, such as developer companies, adopting entities, training database providers, and media platforms or online forums that enable users to share AI-generated content, accounted for less than 5% of incident disclosures. However, this contributes to a blind spot in the understanding of pre-deployment AI incidents and prevents us from developing effective risk prevention strategies during the pre-deployment phase.

Among all incident types we analyzed, the most frequently occurring type is incidents involving `non-consensual imagery, impersonation, fake content` ($n = 78, 39\%$). The commonness of this incident type among all AI privacy and ethical incidents we analyzed raises the question about the protection of people's digital identities and the authenticity of content in the age of AI. Spreading AI-generated deepfakes, manipulated videos, and fabricated content contributes to misinformation, manipulates public opinions and beliefs, and harms individuals and society. These effects are illustrated in Figure 6 and mentioned in Section 4.5, where we found that this incident type often results in damages to individuals or groups ($n = 44, 22\%$ and $n = 28, 14\%$, respectively), or backlash and concerns on AI from the general public ($n = 38, 19\%$). These incident types are often caused by malicious people deliberately abusing AI systems ($n = 39, 19\%$; see Figure 4). Preventing this requires safeguards within AI systems to minimize the potential for misuse, such as watermarking content or embedding detection mechanisms for manipulated media.

More than half ($n = 117, 58\%$) of the incidents in our analysis originated from the organizational decisions made by entities that developed AI systems (AI systems and developers, $n = 67, 33\%$) and the private and public sectors that deployed or maintained these systems (AI adopting organizations and government entities: $n = 58, 29\%$). However, regulations alone may be insufficient to prevent these AI incidents, because both private companies (R5) and public sectors (R6)—who are expected to safeguard citizens' privacy and adhere to ethical standards—were found to use AI against legal regulations (C9) or adopt AI tools known to be problematic (R5-C9 $n = 6, 3\%$ and R6-C9 $n = 6, 3\%$; see Figure 7). Moreover, incidents where AI developers resisted addressing system failures or withheld operational details after incidents (see Section 4.1.2) prove that we need external enforcement and audits to verify that AI developers, adopting organizations, and government entities follow ethical and privacy standards in AI design and deployment.

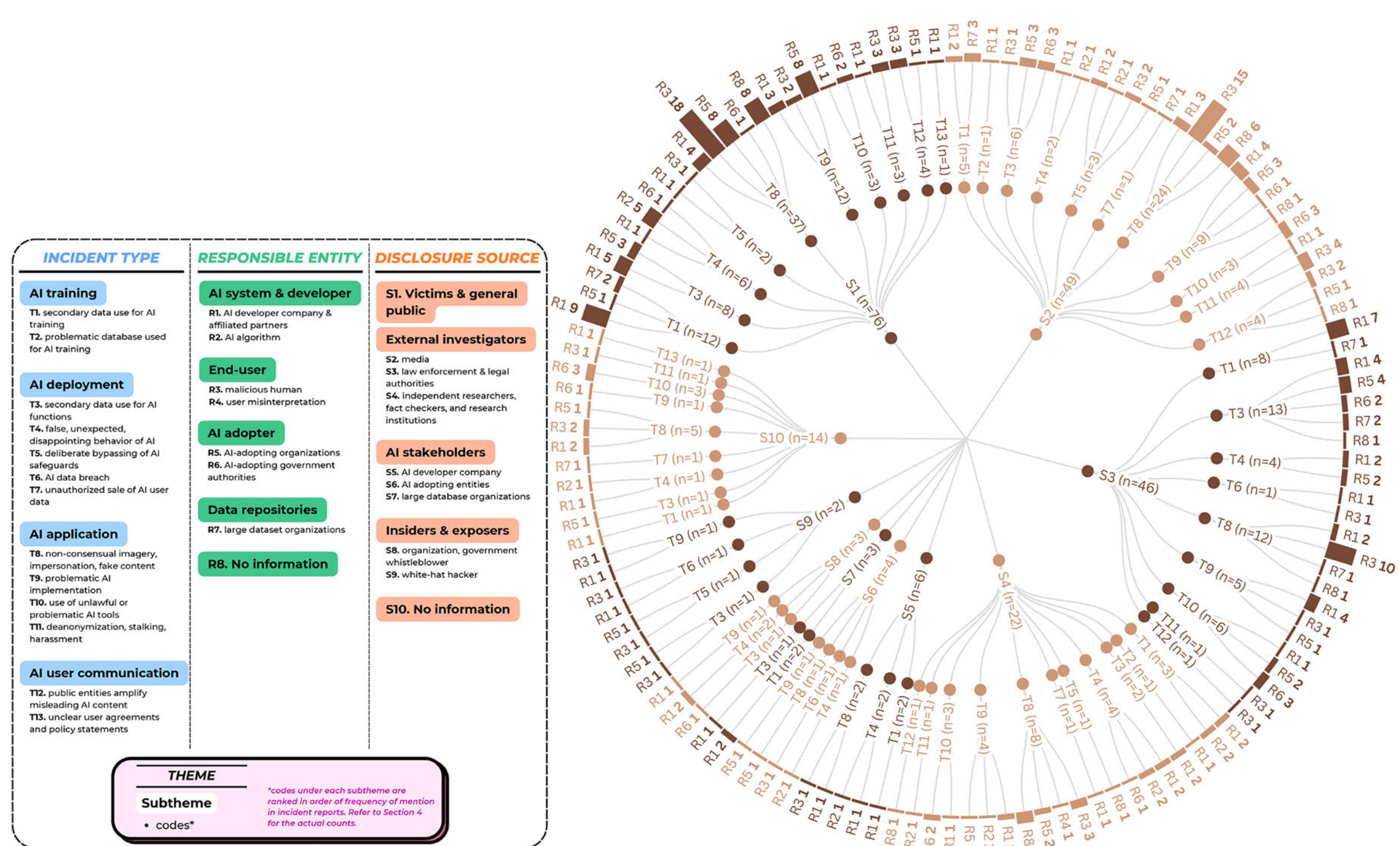

Figure 5. Radial treemap illustrating the relationship between AI incident disclosure source (S), incident types (T), and their corresponding responsible entities (R). Each bar at the outer edge represents the count of incidents attributed to a specific responsible entity within a particular incident type and disclosed by a specific source. The legend on the left presents the name of each disclosure source, incident type, and responsible entity. Longer bars indicate higher frequencies of incidents for a given incident type and responsible entity. For example, we found `victims & general public (S1)` was the source of disclosure for 18 (9%) `non-contextual imagery, impersonation, fake content (T8)` led by `malicious human (R3)`. Incident reports involving multiple disclosure sources, types, or responsible entities are coded with all relevant codes.

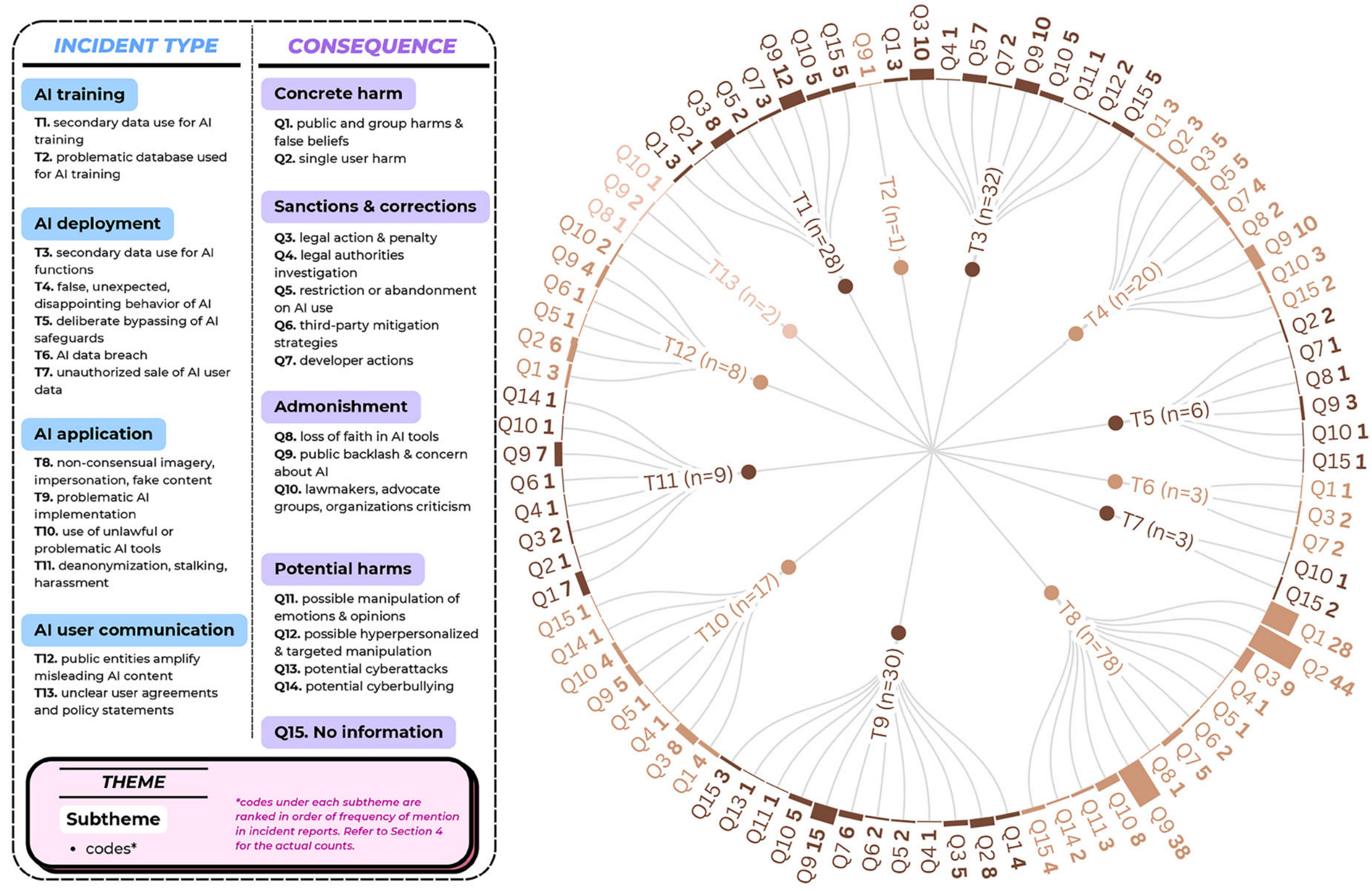

**Figure 6.** Radial treemap illustrating the relationship between AI incident types (T) and their consequential impacts (Q). Each bar at the bars on the outer edge represents the count of incidents led to a particular type of consequence. The legend on the left presents the name of each incident type and consequence. Longer bars indicate higher frequencies of incidents for a given consequence. For example, we found 44 (22%) `non-consensual imagery, impersonation, fake content` has led to `single user harm`. Incident reports involving multiple consequences or types are coded with all relevant codes.

Lastly, as discussed in Section 4.5, the incidents we analyzed most often resulted in reprimands ($n = 111, 55\%$), including public backlash, user concerns, and critiques from lawmakers and ethics experts about AI's societal, ethical, and cultural implications. Additionally, 90 (45%) incidents led to concrete harms, such as privacy loss, erosion of autonomy, reputation damage, financial loss, unequal treatment, loss of online anonymity, emotional distress, physical abuse, and even loss of physical freedom, alongside public panic and the spread of false beliefs. Conversely, only $n = 74$ (37%) incidents resulted in legal penalties to the AI developing or adopting entities, corrective actions and restrictions to the AI systems, or risk-mitigation interventions by third parties. This is concerning. Yet, many incidents that caused public concerns and panic may not strictly violate laws. However, in regions and countries with weak or absent AI privacy and ethical standards and legal frameworks, the exploitation and misuse of AI technologies may hurt people.

### *5.3. Implications for raising public awareness and understanding of AI today*

Prior studies have emphasized the importance of educating users about manipulative tactics (Hadan et al., 2024), where hyper-personalized avatars and realistic ads can blur the line between genuine and distorted content (Buck & McDonnell, 2022; Mhaidli & Schaub, 2021). Our findings indicate that AI can be used to exacerbate the creation of hyperpersonalization and emotional manipulation (e.g., companion). Thus, improving people's AI literacy for them to effectively recognize and resist the AI-driven manipulation and harmful outcomes become essential. Specifically, people need to be made aware of when AI is being used to manipulate emotions or influence decisions through highly targeted, personalized experiences. Literature has started to explore the design of humanistic and ethical AI systems that serve people and address concerns about its potential for exploitation and mislead, rather than replacing them (Capel & Brereton, 2023). Various tools have also been developed to enhance the public's understanding of the potential risks of AI (Bogucka et al., 2024). However, effectively communicating AI risks is challenging, as it requires addressing the diverse experiences, beliefs, and perceptions people have about AI (Bogucka et al., 2024; Long & Magerko, 2020) and tailoring messages to different age groups and educational backgrounds to maximize the impact (Radanliev, 2025). Thus, achieving humanistic and ethical AI design and creating user educational interventions that protect users and enhance their resilience against AI-driven deception requires both academic and industrial sectors' collaboration (Hadan et al., 2024; Mhaidli & Schaub, 2021). Such a collaboration will base educational tools on human psychology and real-world empirical insights to embed ethical guidelines into AI development processes.

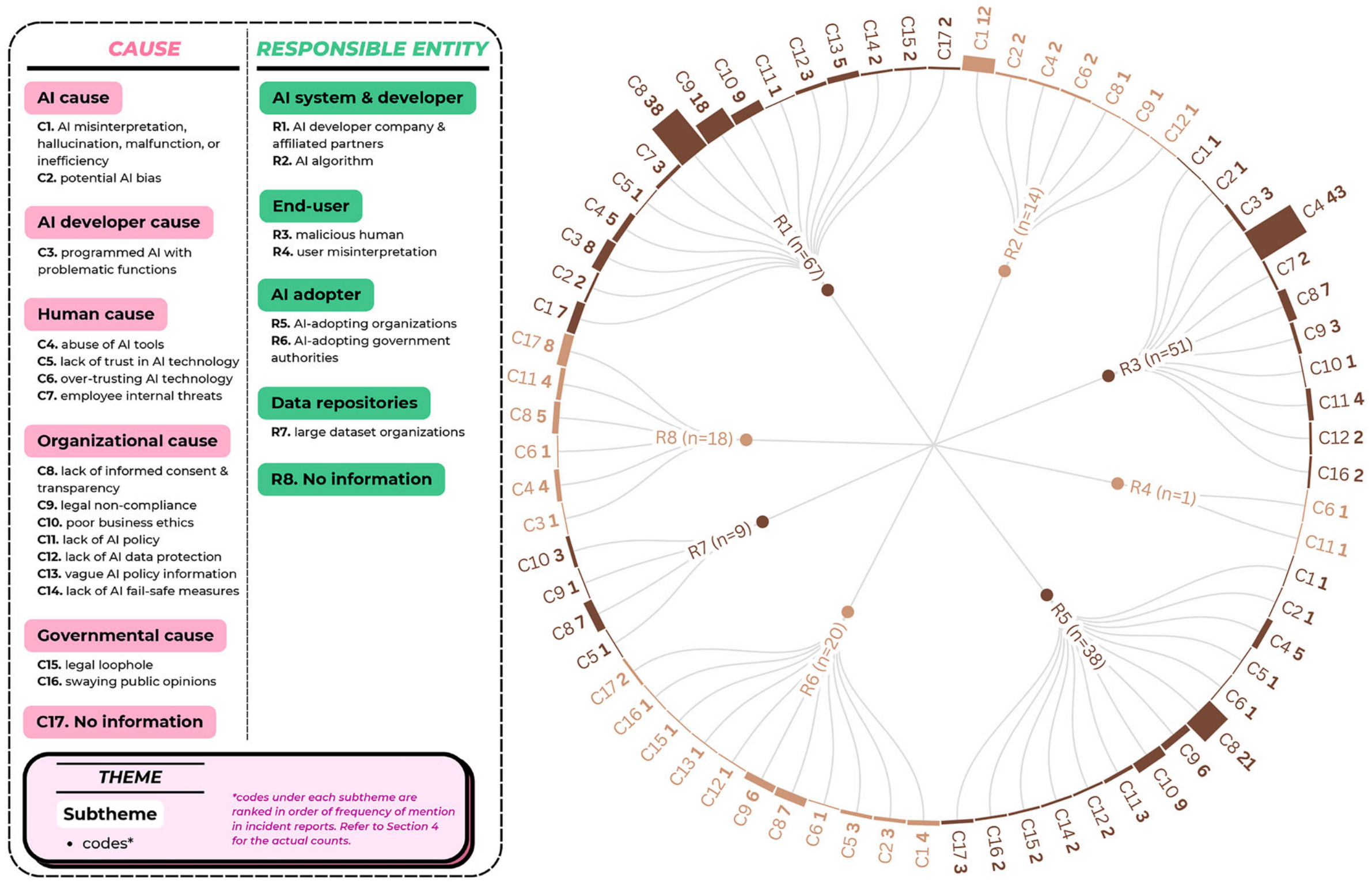

**Figure 7.** Radial treemap illustrating the relationship between AI incident causes (C) and their corresponding responsible entities (R). Each bar at the outer edge represents the count of incidents attributed to a specific cause by a specific responsible entity. The legend on the left presents the name of each cause and responsible entity. Longer bars indicate higher frequencies of incidents for a given cause and responsible entity. For example, we found 38 (14%) incidents due to `lack of informed consent & transparency (C8)` by `AI developer company & affiliated partners (R1)`. Incident reports involving multiple causes or types are coded with all relevant codes.

As AI systems increasingly integrate into our daily lives, it is essential for AI users and the general public to critically evaluate the authenticity and reliability of AI-generated content. While AI optimizes tasks and enhances efficiency (Hadan et al., 2024; Mogavi et al., 2024), our research illustrates the real-world cases of AI in spreading false information, amplifying societal biases, and introducing new forms of discrimination. Alarmingly, we found concerning instances of AI technologies in political and military campaigns to manipulate public opinion, and people's overreliance and blind trust in these systems further exacerbate these issues. The stakes will only grow higher as AI systems become more integrated into decision-making processes across sectors such as healthcare, finance, and law enforcement (Burema et al., 2023). As AI technologies continue to evolve, the consequences of such blind trust will not be limited to minor inconveniences (e.g., being evacuated from retailer stores; AIAAIC Repository, 2024s) but can lead to significant harms, including restrictions on individual autonomy and unjust treatment based on flawed AI interpretations (Slattery et al., 2024; Yampolskiy, 2016). Thus, beyond educational interventions, we suggest that people living in this age of AI need to learn to approach AI-generated content with a critical mindset. For example, when engaging with AI-generated content, it is crucial to verify its accuracy by cross-referencing multiple reputable sources to avoid biases or inaccuracies, rather than blindly trusting it as fact.

### 5.4. Implications for future AI incident reporting

Our findings emphasize the critical need for standardized, comprehensive AI incident reporting frameworks to ensure more consistent and transparent documentation of AI incidents and contextual factors that contributed to the incidents. The lack of specific information in many incident reports in our analysis (see Figure 2 and Appendix E) demonstrates the inadequacy of current practices in AI incident documentation. Although we recognize that some factors might be unknown at the time of reporting, a structured framework can help distinguish between incomplete reports and those that genuinely lack sufficient evidence. Our taxonomy and synthesized contextual factors could serve as a practical foundation, which offers a clear outline and standardized options to streamline and simplify the reporting process for stakeholders.

Our analysis revealed the low incident disclosure rates from AI developers and adopting organizations and government entities, with each contributing less than 5% of the incident reports, similar to the low self-disclosure observed in cybersecurity incident reporting (Hadan et al., 2021; Serrano et al., 2019). It is understandable that the reluctance to self-disclose may originate from a fear of reputational harm, a lack of awareness about privacy and ethical risks, or an underestimation of the value of transparency. However, research from other fields has shown that clear, honest, and active disclosure and communication fosters a positive view, trust, and understanding between users and the developers (Freeman et al., 2022; Radanliev, 2025). Thus, for AI developers and adopting organizations and government entities, publicly reporting incidents is not only a demonstration of accountability but also a means to foster public confidence and promote responsible AI practices (Kulikova et al., 2012; Singer & Cooper, 2009).

To encourage greater disclosure from AI developers and adopting organizations and government entities, one potential solution can be to support anonymous submissions through a confidential database, similar to those successfully used in aviation (The National Aeronautics and Space Administration (NASA), n.d.). This de-identified reporting method can offer a safe space for AI developers and adopting organizations and government entities, and other indirectly involved stakeholders to share detailed accounts without fear of reputation damage (Turri & Dzombak, 2023). Another approach is to mandate AI incident disclosure through regulatory standards, as seen in privacy and cybersecurity frameworks. For instance, the GDPR requires organizations to report certain personal data breaches to supervisory authorities within 72 hr of discovery European Parliament (Council of the European Union, 2018). Similarly, network security standards mandate that responsible entities investigate and report certificate problems within 24 hr (CA/Browser Forum, 2024). We thus recommend implementing similar legal mandates for AI incident disclosure, as it can not only improve incident disclosure rates from AI developers and adopting organizations and government entities but also establish a clear standard for accountability, and ensure that the disclosed information is timely, actionable, and valuable for preventing and mitigating future AI incidents.

### 5.5. Implications for future AI incident detection and prevention

Our analysis revealed that AI developers and adopting organizations and government entities were identified as the primary responsible entities in more than half of incidents we analyzed (see Figure 7 and Appendix E), although the low rate of incident disclosure from these stakeholders has limited our understanding of the specific workflows and contexts in which these incidents occur. These incidents highlight the need for improved methods to detect and prevent AI incidents and for establishing baseline standards for AI development and deployment.

Although rare, we identified five (2%) incidents where disclosures came from whistleblowers within AI developer and adopter organizations, as well as independent white hat hackers. These instances demonstrate the potential of these entities in addressing problems that may be underestimated or deliberately concealed by AI developer and adopter organizations due to business priorities. While researchers have offered guidance and practical tools to help AI developers design responsible and ethical systems (Boyd, 2021; Radanliev, 2025), there is a pressing need for AI governance organizations, as well as AI developers, adopting organizations, and government entities, to implement structured incentives and protections for whistleblowers. These measures are critical for ensuring that harmful AI systems are flagged and addressed, rather than left unchecked in favor of business interests. Fostering such protections promotes a culture of early risk detection, responsible intervention, and ultimately, greater transparency and accountability in AI development and deployment.

Through our analysis, we also identified incidents of malicious exploitation of AI system vulnerabilities, including human abuse of AI systems ($n = 54, 27\%$) and AI data breaches ($n = 3, 1\%$). These findings emphasize the need for robust baseline security controls and safeguards within AI systems and their outputs to prevent misuse. Prior work suggests developers, users, and affected communities should collaborate in integrating transparency, explainability and clarity in AI development (Radanliev, 2025). However, these approaches rely heavily on awareness and ongoing diligence from developers and users. To complement this, we recommend embedding user-facing protections that alert people about AI exploitation at the moment of interaction. For instance, watermarking AI-generated content and embedding detection mechanisms to combat misinformation and deepfakes have been proposed in manipulative design research (Hadan et al., 2024). However, further research is necessary to evaluate their practical implementation and effectiveness in real-world contexts. Future effort is also needed on developing tools to help users identify and critically assess AI-generated content, similar to browser extensions for detecting malicious websites and deceptive design (Schäfer et al., 2024). As AI and immersive technologies evolve, raising awareness of AI-generated content in virtual environments will be crucial, as immersive technologies' realistic simulations and advanced data processing can amplify the impact of hyper-personalization and deepfakes (Hadan, Choong et al., 2024; Hadan, Wang, et al., 2024).

### 5.6. Implications for future AI incident governance and regulation

Our findings suggest the insufficiency of existing organizational and governmental AI governance systems, as over half of the incidents ($n = 117, 58\%$) we analyzed were due to organizational issues among AI developers and adopting organizations and government entities, such as inadequate transparency, failure in obtaining consent, and insufficient AI protection and fail-safe measures. Beyond establishing robust regulations and standards, there was also a need for robust enforcement, as we further observed many incidents resulted from legal non-compliance and unethical practices by both private and governmental entities (see Figure 4). These incidents demonstrate the inability of these entities to consistently, reliably, and voluntarily adhere to privacy and ethical AI practices. Many incident reports in our analysis also contained lawmakers', privacy and ethics advocates', and experts' calls for validation of claims made by AI developers and adopting organizations and government entities regarding the accuracy and safety of their systems (e.g., AIAAIC Repository, 2023n).

Although not a dominant theme in our findings, we identified incidents where children and teenagers were specifically targeted in incidents (see Section 4.1.3). Additionally, we found that a lack of AI governance policies on social media platforms ($n = 12, 6\%$) contributed to the unchecked spread of harmful and falsified content. Many incident reports also documented lawmakers', privacy and ethical advocates and experts' calls for caution regarding the risks synthetic media pose to young audiences, as it can blur the line between reality and fiction, potentially leading to confusion and harm (e.g., AIAAIC Repository, 2023aj). We thus see the urgent need for stricter laws and standards tailored to children-focused AI products, alongside platform-specific AI policies to effectively moderate and mitigate harm, particularly on platforms popular with younger users. For example, age-appropriate design codes, like the UK's Age Appropriate Design Code Office (n.d), establish legal safeguards to protect children online. Regulations like these have the potential to limit AI-driven data collection, prohibit targeted ads for minors, and mandate clear, child-friendly AI content warnings. Additionally, popular platforms can integrate AI filters to detect and block harmful, manipulative, or inappropriate AI-generated content.

Literature highlights that the capacity of AI to aggregate and analyze vast datasets can amplify privacy risks (Lee et al., 2024). Thus, incidents such as AI data breaches ($n = 3, 1\%$), secondary data use for AI training and functions ($n = 28, 14\%$ and $n = 32, 16\%$, respectively), and the unauthorized sale of AI user data ($n = 3, 1\%$) are likely to become increasingly prevalent. Furthermore, in many incident reports, we observed the use of AI in political and military campaigns (e.g., AIAAIC Repository, 2024i, 2024w), and found contributors' concern about the criminalization of AI (e.g., AIAAIC Repository, 2024n, 2023ab) and the threats to election integrity from AI-driven interference (e.g., AIAAIC Repository, 2024w). However, addressing these challenges is complex, because AI applications deemed harmful in some countries may be acceptable—or even encouraged—in others (Yampolskiy, 2016). Given these divergent perspectives, future research should explore how cultural, political, and social values shape regional AI governance. Rather than aiming for full reconciliation, understanding these differences could help policymakers develop AI regulations that align with local priorities while minimizing global risks.

### 5.7. Limitations and future work

While our research offers valuable insights into AI privacy and ethical incidents, we acknowledge several limitations that may guide future work.

First, our analysis focuses exclusively on incidents reported on the AIAAIC repository from 2023 and 2024. As AI technologies evolve, the incidents and the associated contextual factors may change. New risks and consequences may emerge beyond the scope of our dataset. We encourage future researchers to monitor AI incidents longitudinally, identify emerging trends, and address novel harms to inform proactive measures.

Second, our study is limited to publicly disclosed incidents. Thus, incidents that remain undisclosed or unknown to the public are outside our scope. This limitation is reflected in the low disclosure rates from AI developers and adopting organizations and government entities, as well as our taxonomy's inability to fully capture incidents from the internal processes of these entities (Cummings, 2024; Shahriar et al., 2023). We hope our research contributes to developing guidelines that promote systematic reporting of AI-related issues, and encourage more transparent disclosure from developers and adopting organizations and government entities.

Third, our analysis was limited to incident reports explicitly labeled as involving "privacy" and "ethical" issues by contributors to the AIAAIC repository. Although we randomly reviewed 20 unselected reports and found evidence of data saturation in the selected dataset (see Section 3.2), it is still possible that unlabeled yet relevant cases were missed, potentially excluding novel or underexplored issues. Thus, future research can build on our approach by examining a broader range of reports, including those without explicit labels, to capture a more holistic view of AI incidents.

Fourth, our interpretation of incident reports and thematic analysis may reflect biases rooted in our research team's expertise in deceptive design, human-AI interaction, computer privacy, and cybersecurity. To mitigate potential biases, we grounded our taxonomy in established definitions of AI privacy risks and ethical design principles (see Section 3.1). While this approach strengthened the rigor of our analysis, we recognize the importance of future researchers from diverse disciplines replicating our study to offer alternative perspectives and deeper insights.

Lastly, our dataset relies on the ability of contributors to clearly and truthfully describe the incidents. Thus, our analysis may suffer from sampling bias, because certain contextual factors omitted in the reports may have been excluded. Despite this, our sample's size and diversity provide confidence in the robustness of our framework, because it captures a broad range of incidents with recurring contextual factors likely reflected in our analysis.

Notwithstanding these limitations, our study offers a valuable foundation for understanding AI incidents, their contexts, and origins. Our taxonomy and research findings enable future research and provide actionable insights for advancements in AI governance, and the prevention and mitigation of AI-related incidents.

## 6. Conclusion

In this paper, we present a thematic analysis of 202 real-world incidents from 2023 to 2024. This analysis formed the foundation of a comprehensive and empirically grounded taxonomy that categorizes incident types and their contributing factors, such as cause, responsible entities, disclosure sources, and consequences. Our findings expose critical gaps in current AI governance, including insufficient incident reporting, limited regulatory oversight, and a widespread lack of risk mitigation and corrective efforts, especially on social media and among children.

Our research builds on and extends existing work by offering a more comprehensive and detailed AI incident taxonomy. This taxonomy captures cases previously hypothesized in the literature and novel incident types and contributing factors that existing frameworks fail to address. In doing so, we offer a more granular, evidence-based foundation for understanding the causes and consequences of AI incidents for improving incident detection and risk mitigation across the AI lifecycle. From our analysis of real-world AI incidents, we also offer actionable insights that support the development of improved detection systems, proactive governance frameworks, and mandatory incident reporting mechanisms.

As AI continues to shape global policies, economies, and societal norms, the urgency to establish proactive, enforceable governance measures cannot be overstated. Our research calls upon policymakers, industry leaders, and researchers to implement systematic AI risk management strategies that prioritize transparency, public safety, and ethical responsibility. Our findings offer practical value to policymakers, regulators, AI developers, and organizational stakeholders aiming to design and deploy ethical AI systems. We encourage future work to focus on applying and validating these interventions in real-world settings to ensure that AI evolves in ways that are socially responsible, equitable, and safe.

In conclusion, our research establishes a new baseline for AI incident reporting and shapes the way AI failures are studied, discussed, and addressed. As AI's role in society expands, so too must our ability to detect, document, and mitigate its risks. Our taxonomy provides the foundation for doing exactly that.

## Notes

1. AIAAIC Repository. https://www.aiaaic.org/aiaaic-repository.
2. AI Incident Database. https://incidentdatabase.ai/.
3. OECD AI Incidents Monitor (AIM). https://oecd.ai/en/incidents.
4. Dovetail—Qualitative Coding Platform. https://dovetail.com/.
5. Plotly Python Graphing Library. https://plotly.com/python/.
6. Canva. https://www.canva.com/.

## Acknowledgements

We thank the AIAAIC Repository of AI, Algorithmic and Automation Incidents and Controversies (AIAAIC) for serving as a valuable data source in our analysis. Thank you to the editor for the effort in organizing the peer-review process, and thank you to the reviewers for their insightful feedback that improved the quality of this manuscript. We also thank graduate researcher Derrick M. Wang for providing technical support in resolving LaTeX errors.

## Author contributions

CRediT: **Hilda Hadan**: Conceptualization, Data curation, Formal analysis, Investigation, Methodology, Project administration, Resources, Software, Validation, Visualization, Writing – original draft, Writing – review & editing; **Reza Hadi Mogavi**: Conceptualization, Formal analysis, Methodology, Validation, Writing – review & editing; **Leah Zhang-Kennedy**: Funding acquisition, Supervision; **Lennart E. Nacke**: Funding acquisition, Supervision, Writing – review & editing.

## Disclosure statement

The authors declare that there are no conflicts of interest regarding the publication of this article.

## Funding

This research was supported by the Natural Sciences and Engineering Research Council of Canada (NSERC) Discovery Grant (#RGPIN-2022-03353 and #RGPIN-2023-03705), the Social Sciences and Humanities Research Council of Canada (SSHRC) Insight Grant (#435-2022-0476), the Canada Foundation for Innovation (CFI) JELF Grant (#41844). Any opinions, findings, and conclusions or recommendations expressed in this material are those of the author(s) and do not necessarily reflect the views of the NSERC, the CFI, nor the University of Waterloo.

## ORCID

Hilda Hadan http://orcid.org/0000-0002-5911-1405
Reza Hadi Mogavi http://orcid.org/0000-0002-4690-2769
Leah Zhang-Kennedy http://orcid.org/0000-0002-0756-0022
Lennart E. Nacke http://orcid.org/0000-0003-4290-8829

## Data availability statement

The data that support the findings of this study are openly available in the Open Science Framework (OSF) at https://doi.org/10.17605/OSF.IO/SWY5J or https://osf.io/swy5j/.

## About the authors

**Hilda Hadan** is a Postdoctoral Researcher at the Stratford School of Interaction Design and Business, and a member of the HCI Games Group, University of Waterloo. Her research focuses on human factors and ethics in AI systems, deceptive design and player privacy in games and gaming technology.

**Reza Hadi Mogavi** is a Lupina Foundation Postdoctoral Fellow at the University of Waterloo's HCI Games Group and holds a PhD in Computer Science and Engineering from HKUST. His research focuses on responsible AI, immersive interaction, and culturally adaptive technologies, contributing to inclusive and socially responsive approaches in human-centered computing.

**Leah Zhang-Kennedy** is an Associate Professor of Interaction Design and User Experience Research at the Stratford School of Interaction Design and Business, University of Waterloo. Her research focuses on improving safer technology practices, with an emphasis on usable privacy and security, and digital literacy.

**Lennart E. Nacke** is an HCI professor and University Research Chair at the University of Waterloo. He has supported thousands of researchers in improving their writing and reaching greater academic success. He teaches a unique blend of AI tools with time-tested methods and research experiences that improve research strategy.

## Appendix A: Declaration of AI use in manuscript preparation

We acknowledge that we used Grammarly's AI assistant and Typingmind's Claude 3.5 Sonnet AI model for spelling, grammar, punctuation, and clarity editing. The prompt we used was: "make the following sentence [our human-written sentence] more concise and flow better, and fix any grammar errors." Our manuscript was fully verified and edited by our research team. Our decision to use AI tools was guided by the aim of achieving concise and clear academic writing while preserving the "human-touch" in our expression Hadan et al. (2024). Our research team takes full responsibility for the content of the publication. We did not use generative AI for data collection, analysis, or image generation. Figures in this manuscript were created using Python plotly graphing library[5] and pre-built templates on Canva.[6] The Overleaf Gemini AI integration was used to resolve LaTeX errors.

## Appendix B: AIAAIC repository metadata fields overview

**Table B1.** AIAAIC repository field names and definitions.

| Metadata field | Definition* |
|---|---|
| AIAAIC ID | |
| Headline | |
| Type | Each entry is classified as a System, Incident, Issue, or Data. |
| Release | The year (and, on the website, month) a system or dataset/database is soft and/or formally launched. |
| Occurred | The year (and, on the website, month) an incident or issue occurs, or first occurs. |
| Country(ies) | The geographic origin and/or primary extent of the system/incident/issue/data. |
| Sector(s) | The industry (including government and non-profit) sector primarily targeted by the system or adversarial attack. |
| Deployer(s) | The name of the individual(s), group(s), or organisation(s) deploying/managing the system or dataset/database involved in an incident or issue on a day-to-day basis, or the platforms on which the system is hosted or being carried. |
| Developer(s) | The name of the individual(s) or organization(s) involved in designing, or developing/providing the system or dataset/database, and/or that commissions a system to be developed with a view to placing it on the market or putting it into service under its own name or trademark whether for payment or free of charge or that adapts general purpose systems for a specific intended purpose. There may be multiple providers along the system lifecycle. |
| System name(s) | The name of the system, set of systems, or dataset/database involved in an incident or issue. |
| Technology(ies) | The type(s) of technology deployed in the system. |
| Purpose | The aim(s) of the system or dataset/database. |
| Media trigger(s) | The internal or external trigger for a public issue or incident. |
| Issue(s) | The issues of concern posed by a system, its governance and/or technology, or by third-parties. |
| Harm(s) | Harms are the actual negative impacts caused by an incident, system, or dataset/database. The harms may be caused directly (sometimes known as 'first-level' harms) or indirectly ('second-level') harms. |

*Note:* *Definitions retrieved on September 24, 2024, upon our data collection, from the AIAAIC website https://www.aiaaic.org/aiaaic-repository/classifications-and-definitions.

## Appendix C: Taxonomy of 12 AI-related privacy risks

The AI privacy risks presented in Table C2, adapted from Lee et al. (2024), defined what we considered privacy violations during the screening and coding of incident reports. See Section 3.1.

**Table C2.** The 12 AI-specific privacy risks Lee et al. (2024) Used for our data screening and analysis.

| Privacy risks | Description Lee et al. (2024) |
|---|---|
| Surveillance | The monitoring or recording of people's activities, with AI intensifying these risks by enabling large-scale, pervasive data collection. |
| Identification | Linking specific data points to a person's identity, with AI amplifying these risks due to its scale, speed, persistence, and widespread application. |
| Aggregation | The process of combining multiple data points about a person to infer information beyond what is explicitly provided, with AI intensifying these risks through its scale, speed, ubiquity, and predictive capabilities. |
| Phrenology/Physiognomy | The practice of inferring personal, social, or emotional traits from an individual's physical features, a risk AI amplifies by learning and acting on spurious correlations between such features and sensitive attributes. |
| Secondary use | The use of personal data for purposes beyond the original intent without user consent, a risk intensified by AI's ability to repurpose data and generate new capabilities or models from existing datasets. |
| Exclusion | The lack of transparency and user control over data usage, with AI amplifying this risk by leveraging personal data for training without individuals' knowledge or consent. |
| Insecurity | Inadequate protection of personal data from leaks or unauthorized access, with AI heightening this risk by introducing new vulnerabilities through its integration and data pipelines. |
| Exposure | The revealing of highly sensitive or concealed personal information, with AI intensifying this risk through generative techniques and the inference of private data, preferences, and intentions. |
| Distortion | The dissemination of false or misleading information about people, with AI amplifying this risk through the creation of highly realistic fake images and audio that are difficult for people to detect as inauthentic. |
| Disclosure | The improper revelation or sharing of people's data, with AI exacerbating this risk by inferring hidden information from raw data and by using personal data to train models. |
| Increased Accessibility | The increased ease with which sensitive personal information can be accessed by a broader audience, a risk amplified by AI through the public release of large-scale datasets used to train and improve models. |
| Intrusion | The disturbances of a person's physical solitude, with AI amplifying this risk by enabling centralized and pervasive surveillance infrastructures. |

## Appendix D: Overview of 11 core AI ethics principles

The AI ethics principles presented in Table D3, adapted from Jobin et al. (2019), defined what we considered ethical violations during the screening and coding of incident reports. See Section 3.1.

**Table D3.** The 11 AI ethics principles Jobin et al. (2019) Used for our data screening and analysis.

| Principle | Description Jobin et al. (2019) |
|---|---|
| Transparency | The efforts that enhance the explainability, interpretability, and disclosure of information regarding AI systems, including their data use, decision-making processes, limitations, and impacts. |
| Justice and fairness | The prevention and mitigation of bias, discrimination, and inequality, promoting inclusion, diversity, and equal access to AI systems, data, and benefits. |
| Non-maleficence | The avoidance of foreseeable or unintentional harm, including risks such as discrimination, privacy violations, and physical or psychological injury. |
| Responsibility and accountability | The clear attribution of obligations and liability for AI systems' actions and impacts, emphasizing integrity, transparency, and the need for remedy when harm occurs. |
| Privacy | The support of technical measures (e.g., privacy by design, data minimization), legal compliance, and increased awareness to protect people's personal data through data protection, security, and responsible handling practices. |
| Beneficence | The promotion of well-being, flourishing, and the common good through AI systems design and deployment, ensuring that AI benefits humanity broadly across individuals, society, and the environment. |
| Freedom and autonomy. | The assurance of people's rights to make informed, independent decisions and maintain control over their personal data and digital experiences. |
| Trust | The confidence that users, developers, and society place in AI systems, organizations, and design principles to act reliably, transparently, and in alignment with shared values and public expectations. |
| Sustainability | The design, development, and deployment of AI systems in ways that protect the environment, promote social equity, and ensure long-term ecological and societal well-being. |
| Dignity | The respect, preservation, and promotion of human worth and rights, ensuring that AI systems do not harm, coerce, classify, or interact with individuals in ways that compromise their autonomy or humanity. |
| Solidarity | The fair distribution of AI's benefits, the protection of vulnerable groups, and the reinforcement of social cohesion through supportive systems like social safety nets. |

## Appendix E: Descriptive statistics of our AI privacy and ethical incident taxonomy

**Table E4.** Taxonomy of AI incidents' type, source of disclosure, cause, responsible entity, and consequence (*$N = 202$).

| Theme/subtheme | Code | n (%)* |
|---|---|---|
| *Theme: Incident Type* | | |
| AI training ($n = 29, 14\%$) | secondary data use for AI training | 28(14%) |
| | problematic database used for AI training | 1 (<1%) |
| AI deployment ($n = 62, 31\%$) | secondary data use for AI functions | 32 (16%) |
| | AI false, unexpected, disappointing behavior | 20 (10%) |
| | deliberate bypassing of AI safeguards | 6 (3%) |
| | AI data breach | 3 (1%) |
| | unauthorized sale of user data | 3 (1%) |
| AI application ($n = 124, 61\%$) | non-consensual imagery, impersonation, fake content | 78 (39%) |
| | problematic AI implementation | 30 (15%) |
| | use of unlawful/problematic AI tools | 17 (8%) |
| | deanonymization, stalking, harassment | 9 (4%) |
| AI user communication ($n = 10, 5\%$) | public entity amplified of misleading content | 8 (4%) |
| | unclear user agreements and policy statements | 2 (1%) |
| *Theme: Cause* | | |
| no information | – | 15 (7%) |
| AI cause ($n = 25, 12\%$) | AI misinterpretation, hallucinations, malfunctions, inefficiency | 21 (10%) |
| | potential AI bias | 7 (3%) |
| AI developer cause | programmed AI with problematic functions | 10 (5%) |
| Human cause ($n = 69, 34\%$) | abuse of AI tools | 54 (27%) |
| | lack of trust on AI | 6 (3%) |
| | over-trusting AI | 5 (2%) |
| | employee internal threats | 4 (2%) |
| Organizational cause ($n = 117, 58\%$) | lack of informed consent & transparency | 81 (40%) |
| | legal non-compliance | 32 (16%) |
| | poor business ethics | 22 (11%) |
| | lack of AI policy | 12 (6%) |
| | lack of data protection | 6 (3%) |
| | vague policy information | 6 (3%) |
| | lack of AI fail-safe measures | 4 (2%) |
| Governmental cause ($n = 8, 4\%$) | legal loophole | 5 (2%) |
| | swaying public opinions | 3 (1%) |
| *Theme: Responsible Entity* | | |
| no information | – | 18 (9%) |
| AI systems and developers ($n = 77, 38\%$) | AI algorithm | 14 (7%) |
| | AI developer company, affiliated partners | 67 (33%) |
| End-users ($n = 52, 26\%$) | malicious human | 51 (25%) |
| | user misinterpretation | 1 (<1%) |
| AI adopters ($n = 58, 29\%$) | AI-adopting organization | 38 (19%) |
| | AI-adopting government authorities | 20 (10%) |
| Data repositories | large dataset organizations | 9 (4%) |
| *Theme: Source of Disclosure* | | |
| no information | – | 14 (7%) |
| victims and the general public | – | 76 (38%) |
| external investigators and authorities ($n = 108, 53\%$) | media | 49 (24%) |
| | law enforcement, legal authorities | 46 (23%) |
| | researchers, research institutions, fact checkers | 22 (11%) |
| AI development and application stakeholders ($n = 13, 6\%$) | AI developer company | 6 (3%) |
| | AI adopting entities | 4 (2%) |
| | large database organizations | 3 (1%) |
| insiders and exposers ($n = 5, 2\%$) | organization/government whistleblower | 3 (1%) |
| | white-hat hacker | 2 (1%) |
| *Theme: Consequence* | | |
| | no information | 18 (9%) |
| Concrete harm ($n = 90, 45\%$) | public/group harms, false beliefs | 44 (22%) |
| | single user harm | 56 (28%) |
| Punitive actions or corrections ($n = 74, 37\%$) | legal action, penalty | 42 (21%) |
| | legal authorities investigation | 3 (1%) |
| | restriction or abandonment on AI use | 17 (8%) |
| | third-party mitigation strategies | 4 (2%) |
| | developer actions | 20 (10%) |
| Admonishment ($n = 111, 55\%$) | loss of faith in AI tools | 4 (2%) |
| | public user backlash, concern about AI | 91 (45%) |
| | lawmakers, advocate groups, organizations criticism | 29 (14%) |
| Potential harms ($n = 10, 5\%$) | possible emotional, opinions manipulation | 5 (2%) |
| | possible hyperpersonalized & targeted manipulation | 2 (1%) |
| | potential cyberattacks | 1 (0%) |
| | potential cyberbullying | 3 (1%) |

These categories do not mutually exclusive. We include a detailed overview in F, including descriptions and examples.

## Appendix F: Detailed overview of our AI privacy and ethical incident taxonomy

**Table F5.** Taxonomy of AI incidents' type, source of disclosure, cause, responsible entity, and consequence.

| Code | Example AI incident report |
| --- | --- |
| **Theme: Incident Type** | |
| **T1. Secondary data use for AI training:** the repurposing of user data to train AI models | LinkedIn scrapes users data without informing or gaining their consent to train its own AI models (AIAAIC Repository, 2024a). |
| **T2. Problematic database used for AI training:** the development of an AI system using datasets that are biased, inaccurate, or unrepresentative, leading to AI systems that perpetuate discrimination, generate biased outputs, or make unreliable decisions. | The dataset used to train Google and Meta's large language models included racist, pornographic, and copyright-protected content, introducing offensive, false, and biased material into the AI models (AIAAIC Repository, 2023b). |
| **T3. Secondary data use for AI functions:** the repurposing of user data to power AI-driven features or services. | A government agency uses real citizens' personal data to quietly test AI-powered analytics software that enables cross-jurisdictional data sharing and analysis (AIAAIC Repository, 2023d). |
| **T4. AI false, unexpected, disappointing behavior:** an AI system behaves in ways that deviate from its intended functionality, produces incorrect or unreliable outputs, or fails to meet user expectations even when operating as designed. | An AI chatbot falsely accused a German journalist of serious crimes due to AI's misinterpretation of the journalist's extensive career on court cases involving abuse and fraud (AIAAIC Repository, 2024d). |
| **T5. Deliberate bypassing of AI safeguards**: individuals or groups exploit AI vulnerabilities, manipulate AI systems, or override built-in safety mechanisms or user agreements within AI tools to achieve certain objectives. | A person used prompt injection to manipulate AI chatbot to expose users' precise location data, despite its built-in restrictions on the disclosure of such information (AIAAIC Repository, 2023g). |
| **T6. AI data breach:** the data used or exposed by an AI system is compromised through unauthorized access, leakage, or exploitation due to vulnerabilities in its algorithms, data storage, or operational infrastructure. | A group of hackers breached the AI hiring chatbot, gained access to sensitive information about job applicants, fast-food franchises, and the company itself (AIAAIC Repository, 2024f). |
| **T7. Unauthorized sale of AI user data:** the monetization of AI user data. | An AI chatbot storing user photos, videos, and conversations sells user data to advertisers (AIAAIC Repository, 2023i). |
| **T8. Non-consensual imagery, impersonation, fake content:** the use of AI tools, such as deepfake technology, to create hyper-realistic images, audio, or videos portraying individuals in situations they never participated in. | A travel agency used AI tools to create a deepfake likeness of a professional model for an advertisement without her consent (AIAAIC Repository, 2024ab). |
| **T9. Problematic AI implementation:** the deployment of AI tools in ways that lead to unintended harms, privacy and ethical concerns, or societal backlash. | Microsoft Recall, a part of Windows 11 system AI feature intended to help users find they've seen on their PC, raised privacy issues as it takes screenshots of a user's screen every few seconds (AIAAIC Repository, 2024l). |
| **T10. Use of unlawful/problematic AI tools:** the deployment of AI systems that are known to be controversial, unethical, or in violation of legal and regulatory standards. | Canadian Tire was found by British Columbia's privacy commissioner for illegally using AI-powered facial recognition technology in its stores to collect customers' images and videos (AIAAIC Repository, 2023ag). |
| **T11. Deanonymization, stalking, harassment:** the use of AI tools to re-identify individuals from anonymized data, such as linking publicly shared images or datasets back to a person's real identity. | A 'digital peeping Tom' used a facial recognition platform to identify the real identities of anonymous porn stars by uploading screenshots from adult films and tracking their online presence (AIAAIC Repository, 2023l). |
| **T12. Public entity amplify misleading AI content:** public figures inadvertently or deliberately disseminate AI-generated articles, images, videos, or social media posts that misinform the public, distort facts, or spread propaganda. | Tesla CEO Elon Musk shared a video featuring an AI-generated voice clone falsely portraying U.S. Vice President Kamala Harris making statements she never actually said (AIAAIC Repository, 2024w). |
| **T13. Unclear user agreements and policy statements:** the terms of service, privacy policies, and consent mechanisms of AI systems are complex, vague, and difficult for users to understand the implications of their engagement. | Adobe's updated terms of service, allowing their use of user content for improving automated services, were found to be overly broad and vague, raising concerns about potential access to sensitive projects (AIAAIC Repository, 2024r). |
| **Theme: Cause** | |
| **C1. AI misinterpretation, hallucinations, malfunctions, inefficiency:** the incident occurred because of errors and limitations in an AI system's performance that lead to incorrect, unreliable, or unintended outputs. | An AI chatbot falsely accused a German journalist of serious crimes due to AI's misinterpretation of the journalist's extensive career on court cases involving abuse and fraud (AIAAIC Repository, 2024d). |
| **C2. Potential AI bias (racism/inequality):** the incident occurred because AI systems produce discriminatory outcomes or reinforce societal inequalities due to biased training data, flawed algorithms, or insufficient consideration of diverse perspectives during development. | The facial recognition system in a supermarket falsely identified a Maori woman as a shoplifter by mismatching her with another Maori woman's photo, leading to her being racially discriminated against and publicly embarrassed (AIAAIC Repository, 2024s). |
| **C3. Programmed AI with problematic functions:** the incident occurred because the AI system is intentionally designed with unethical functionalities that can lead to harmful outcomes. | An AI-powered software allows employers to monitor workers' activities, locations, and performance data, predict task durations, evaluate individual performance, leading to reduced worker autonomy and increased work stress (AIAAIC Repository, 2024p). |
| **C4. Abuse of AI tools:** the incident occurred because of intentional use or exploitation of AI systems by individuals or groups to achieve unethical, illegal, or harmful objectives. | A man was arrested for stalking, doxing, and harassing a female professor, using AI tools to generate fake nude images and creating a chatbot in her likeness to share her personal information online (AIAAIC Repository, 2024v). |

**Table F5.** Continued.

| Code | Example AI incident report |
|---|---|
| **C5. Lack of trust in AI technology:** the incident occurred because of the skepticism or fear surrounding AI technology, leading to legal complaints that opposed or delayed the adoption of AI systems. | South Korea's plan to introduce AI-powered digital textbooks for personalized learning has been delayed after opposition from over 56,000 parents, citing concerns about children's development and well-being (AIAAIC Repository, 2024o). |
| **C6. Over-trusting AI technology:** the incident occurred because individuals or organizations rely blindly on the decisions, predictions, or recommendations made by AI systems without adequately verifying their accuracy. | A woman was misidentified by a facial recognition system at a Foodstuffs supermarket as a banned shoplifter; despite providing three forms of photo identification to prove her innocence, staff accused her of theft and demanded she leave AIAAIC Repository (2024s). |
| **C7. Employee internal threats:** the incident occurred because of the malicious actions by employees within AI developer or adopter companies. | Employees at Amazon Ring had unrestricted access to female users' private videos recorded in bedrooms and bathrooms, enabling them to view, download, and use the footage however they liked (AIAAIC Repository, 2023y). |
| **C8. Lack of informed consent, transparency:** the incident occurred because users were not adequately informed about how their data is collected, processed, and used by the AI system. | Meta used public photos and posts from Australian Facebook and Instagram to train its generative AI models without notifying users or obtaining their consent (AIAAIC Repository, 2024b). |
| **C9. Legal non-compliance:** the incident occurred because the AI system was deployed in a manner that breaches established legal standards. | The Dutch regulator fined Clearview AI for failing to comply with data access requests and for processing the biometric data of Dutch citizens without a legal basis (AIAAIC Repository, 2024z). |
| **C10. Poor business ethics:** deliberate deployment of untested, unsafe, or unethical AI systems for purposes such as manipulating users or spreading disinformation, disregarding societal and environmental impacts. | A Tesla whistleblower alleged that the company prioritized business growth over safety, continuing to market and deploy its self-driving technology despite being aware of safety risks in its Autopilot system (AIAAIC Repository, 2023n). |
| **C11. Lack of AI policy:** the incident occurred because of the absence of effective guidelines, governance, or monitoring mechanisms on digital platforms to prevent the spread of harmful, fake, or inaccurate AI-generated content. | A widespread deepfake video scam on TikTok exposed TikTok's failure to moderate and prevent the spread of harmful AI content (AIAAIC Repository, 2023z). |
| **C12. Lack of data protection:** the incident occurred because of failures in implementing adequate security measures to protect AI systems and the sensitive data they handle. | An AI chatbot storing user photos, videos, and conversations fails to protect user data due to weak password requirements and lack of age verification (AIAAIC Repository, 2023h). |
| **Theme: Cause** | |
| **C13. Vague policy information:** the incident occurred because of the absence of clear, detailed, and transparent policy statements outlining how the AI system operate, use of user data, be governed, or address potential risks. | Adobe's updated terms of service, allowing their use of user content for improving automated services, were found to be overly broad and vague, raising concerns about potential access to sensitive projects (AIAAIC Repository, 2024r). |
| **C14. Lack of AI fail-safe measures:** the incident occurred because of the absence of strategies and protocols designed to prevent or mitigate the harmful outcomes arising from AI system malfunctions and unexpected behaviors. | The US retailer Rite Aid failed to implement reasonable precautions in deploying facial recognition technology, leading to false accusations of theft against innocent customers (AIAAIC Repository, 2023t). |
| **C15. Swaying public opinions:** the incident occurred because of the deliberate use of AI systems to influence, exploit, or alter people's thoughts, emotions, decisions, or behaviors. | AI tools were used to create deepfake videos and fabricated news segments to manipulate public opinion during the election (AIAAIC Repository, 2023ac). |
| **C16. Legal loophole:** the incident occurred because of the gaps in legal frameworks allowing AI systems to be deployed without oversight, accountability, or safeguards for bias, privacy, security, and ethics. | Many legal jurisdictions fail to protect people from deepfakes generated by AI-powered bots on Telegram due to the absence of specific regulations (AIAAIC Repository, 2024af). |
| **C17. No information:** the cause was not described in the incident report. | e.g., AIAAIC Repository (2023b, 2024ai). |
| **Theme: Consequence** | |
| **Q1. Public and group harms & false beliefs:** the societal or collective damage from the use of AI or dissemination of inaccurate AI outcomes. | Employees at Amazon Ring had unrestricted access to female users' private videos recorded in bedrooms and bathrooms, enabling them to view, download, and use the footage however they liked (AIAAIC Repository, 2023y). |
| **Q2. Single user harm:** negative impacts experienced by an individual, including biased or discriminatory decisions, privacy breaches, exposure to misinformation, financial losses, or emotional distress. | An AI chatbot falsely accused a German journalist of serious crimes due to AI's misinterpretation of the journalist's extensive career on court cases involving abuse and fraud (AIAAIC Repository, 2024d). |
| **Q3. Legal action & penalty:** the enforcement of legal measures, such as lawsuits, fines, regulatory penalties, or sanctions, imposed on individuals, organizations, or entities responsible for the AI incident. | Meta (formerly Facebook) faced legal action from the Texas Attorney General for illegally collecting biometric data with facial recognition (AIAAIC Repository, 2024ad). |
| **Q4. Legal authorities investigation:** the regulatory bodies, law enforcement, or other legal entities announce or suggest the necessity of an official inquiry into the AI system or its developers/operators. | An advertisement by Volkswagen Brazil using AI and deepfake technology to simulate a duet between the late Brazilian singer and her daughter has prompted an investigation by Brazil's advertising regulatory authority into potential ethical breaches (AIAAIC Repository (2023aj). |
| **Q5. Restriction or abandonment of AI tools:** the discontinuation, restriction, or outright prohibition of the AI systems. | South Korea's plan to introduce AI-powered digital textbooks for personalized learning has been delayed after opposition from over 56,000 parents (AIAAIC Repository, 2024o). |
| **Q6. Third-party mitigation strategies:** the interventions by independent organizations, regulators, advocacy groups, or other entities external to the developer, AI adopter, or affected individuals to address, mitigate, or resolve harm caused by an AI incident. | A website was launched as a third-party effort to mitigate harm caused by deepfake pornography, providing resources and support to victims after high school students used deepfake technology to create and share nude images of female classmates (AIAAIC Repository, 2023j). |

**Table F5.** Continued.

| Code | Example AI incident report |
|---|---|
| **Q7. Developer actions:** the AI developers take actions to implement corrective or protective measures in response to the AI incident. | In response to GPT-4o model's unintentional imitation of users' voices during testing, OpenAI implemented measures to detect and block deviate outputs, and preset voice designs using voice actors (AIAAIC Repository, 2024ah). |
| **Q8. Loss of faith in AI tools:** a decline in public trust and confidence in the reliability of AI systems and their developer and adopter companies. | Several high-profile universities have opted to disable Turnitin's AI writing detection tool, citing concerns over its accuracy and the risk of falsely accusing students of academic dishonesty (AIAAIC Repository, 2023r). |
| **Theme: Consequence** | |
| **Q9. Public user backlash & concern about AI:** the public criticism, protests, boycotts, or declining usage of AI products and services from users or general public due to dissatisfaction, fear, or mistrust. | A supermarket's use of an AI-driven dynamic pricing system sparked fears of corporate profiteering, potential privacy violations, and disproportionate impacts on low-income customers due to price surges (AIAAIC Repository, 2024ai). |
| **Q10. Lawmakers, advocates groups, organizations criticism:** the negative reactions, scrutiny, or public condemnation from government officials, advocacy groups, industry watchdogs, or independent organizations. | Lawmakers criticized the Worldcoin for operating without proper regulation, collecting excessive biometric data from children, violating Kenyan law, and potential espionage (AIAAIC Repository, 2023ab). |
| **Q11. Possible manipulation of emotional & opinions:** concerns or speculation that AI systems may create opportunities for influencing people's emotions, beliefs, or decisions in ways that could be unethical or harmful. | AI companion apps, marketed as sources of emotional support for lonely people, could foster deep emotional connections and encourage users to share intimate details, leading to dependency and unrealistic expectations (AIAAIC Repository, 2023h). |
| **Q12. Possible hyper-personalized & targeted manipulation:** concerns that AI systems could exploit user data to deliver highly tailored content designed to influence individuals' decisions, beliefs, or behaviors in unethical or harmful ways. | Vending machines with facial recognition and "demographic sensors" were analyzing people's age and gender to make AI-powered product recommendations (AIAAIC Repository, 2024aa). |
| **Q13. Potential cyberattacks:** concerns that AI systems could be exploited to facilitate or enhance cyberattacks. | Researchers discovered that six commercial AI tools can be manipulated to generate code for breaching systems, stealing sensitive data, tampering with databases, or launching denial-of-service attacks (AIAAIC Repository, 2023al). |
| **Q14. Potential cyberbully:** concerns that AI technologies could enable or amplify cyberbullying by making it easier for people to create and disseminate harmful, targeted, or abusive content. | A man was arrested for stalking, doxing, and harassing a female professor, using AI tools to generate fake nude images and creating a chatbot in her likeness to share her personal information online (AIAAIC Repository, 2024v). |
| **Q15. No information:** the consequence from the incident was not described. | e.g., AIAAIC Repository (2023am). |
| **Theme: Source of Disclosure** | |
| **S1. Victims & general public:** the incident was disclosed by general public, everyday users, or affected people. | A German journalist discovered an AI chatbot falsely accused him of serious crimes, as it misinterpreted his extensive reporting on court cases involving abuse and fraud (AIAAIC Repository, 2024d). |
| **S2. Media:** the incident was disclosed by journalists, news outlets, or other media organizations. | A media investigation revealed that London police officers used a controversial facial recognition service, violating their own stated policies and procedures (AIAAIC Repository, 2024ac). |
| **S3. Law enforcement & legal authorities:** the incident was disclosed by legal authorities through investigations, legal proceedings, law enforcement. | Brazil's advertising regulatory authority has launched an investigation into potential ethical breaches involving the use of AI deepfake technology in an advertisement that simulated a duet between a late Brazilian singer and her daughter (AIAAIC Repository, 2023aj). |
| **S4. Independent researchers, fact checkers, and research institutions:** the incident was disclosed by academically or professionally driven individuals or organizations through rigorous methods, technical analysis, and evidence-based research. | Researchers discovered that six commercial AI tools can be manipulated to generate code for breaching systems, stealing sensitive data, tampering with databases, or launching denial-of-service attacks (AIAAIC Repository, 2023al). |
| **S5. AI developer company:** the incident was disclosed voluntarily by the organization responsible for creating, deploying, or managing the AI system. | OpenAI publicly disclosed GPT-4o model's unintentional imitation of users' voices during testing, and actively implemented safety measures (AIAAIC Repository, 2024ah). |
| **Theme: Source of Disclosure** | |
| **S6. AI adopting entities:** the incident was disclosed by the entity using the AI system, such as a corporation, institution, or public sector body. | A facial recognition company reported unauthorized access to its client login system and is actively cooperating with law enforcement and federal agencies in the ongoing investigation (AIAAIC Repository, 2024x). |
| **S7. Large database organizations:** the incident was disclosed by the data repositories that aggregate, maintain large datasets. | Reddit warned AI companies against scraping data from its platform without permission (AIAAIC Repository, 2024y). |
| **S8. Organization, government whistleblower:** the incident was disclosed by an individual with insider knowledge or technical expertise from the AI developer or adopter organizations. | A Tesla whistleblower alleged that the company prioritized business growth over safety, continuing to market and deploy its self-driving technology despite being aware of safety risks in its Autopilot system (AIAAIC Repository, 2023n). |
| **S9. White-hat hacker** | |
| **No information:** the source entity that disclosed the incident is not described. | |

(*continued*)

**Table F5.** Continued.

| Code | Example AI incident report |
|---|---|
| **Theme: Responsible Entity** | |
| **R1: AI developer company & affiliated partners:** the organizations and collaborating entities responsible for designing, developing, deploying, or maintaining an AI system. | An AI-powered software was designed to allow employers to monitor workers, leading to reduced worker autonomy and increased work stress (AIAAIC Repository, 2024p). |
| **R2. AI algorithm:** the AI systems' the underlying computational model or set of rules that process input data to generate predictions, decisions, or outputs. | An AI chatbot falsely accused a German journalist of serious crimes due to AI's misinterpretation of the journalist's extensive career on court cases involving abuse and fraud (AIAAIC Repository, 2024d). |
| **R3. Malicious human:** an individual or group of individuals who intentionally exploit, manipulate, or misuse artificial intelligence systems. | A man was arrested for stalking, doxing, and harassing a female professor, using AI tools to generate fake nude images and creating a chatbot in her likeness to share her personal information online (AIAAIC Repository, 2024v). |
| **R4. User misinterpretation:** an individual or group of individuals who falsely believed the AI system's outcomes. | A deepfake video promoting the use of "vegan grenades" was mistakenly believed to be real, despite being explicitly labeled as satire (AIAAIC Repository, 2023af). |
| **R5. AI-adopting organizations:** the private sectors that integrate AI systems into its operations as tools or services. | A supermarket uses AI-powered dynamic pricing system to adjust product prices in real time based on factors like demand and customer data (AIAAIC Repository, 2024ai). |
| **R6. AI-adopting government authorities:** the public sectors that implement AI systems to assist with decision-making, manage services, or support public programs. | South Korea's plan to introduce AI-powered digital textbooks for personalized learning has been delayed after opposition from over 56,000 parents (AIAAIC Repository, 2024o). |
| **R7. Large dataset organizations:** the data repositories that aggregate, maintain large datasets used for training AI systems. | A web browser company was accused of selling user data, including copyrighted content, without user consent to third parties for AI training (AIAAIC Repository, 2023i). |
| **R8. No information:** the entity responsible for the incident is not described. | |

We note that these categories do not mutually exclusive.